%% file: main.tex
\documentclass{ieeeaccess}
\usepackage{cite}
\usepackage{amsmath,amssymb,amsfonts}
\usepackage{algorithmic}
\usepackage{graphicx}
\usepackage{textcomp}
\usepackage{hyperref}
\usepackage{braket}
\usepackage{booktabs}
\usepackage{algorithm2e}

\newcommand{\orcidlink}[1]{\,\raisebox{0.2em}{\href{https://orcid.org/#1}{\includegraphics[height=0.8em]{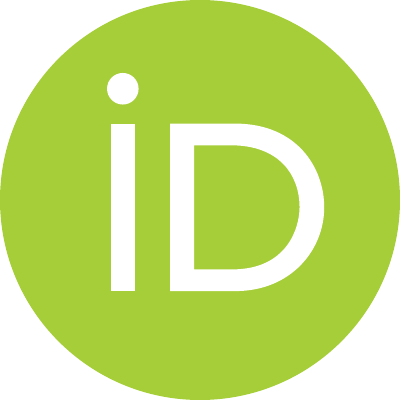}}}}
\newcommand{\coo}{CO\textsubscript{2}}
\DeclareMathOperator{\sign}{sign}
\DeclareMathOperator*{\argmin}{arg\,min}

\def\BibTeX{{\rm B\kern-.05em{\sc i\kern-.025em b}\kern-.08em
    T\kern-.1667em\lower.7ex\hbox{E}\kern-.125emX}}

\newcommand{\newp}[2][]{#2}

\def\linkurl#1{\url{#1}}

\renewcommand{\emph}{\textit}

\newcommand{\zm}{z_\text{max}}

\begin{document}
\history{Date of publication xxxx 00, 0000, date of current version xxxx 00, 0000.}
\doi{10.1109/TQE.2020.DOI}

\title{Incentivising Demand Side Response through Discount Scheduling using Hybrid Quantum Optimization}
\author{\uppercase{%
David Bucher\authorrefmark{1}\orcidlink{0009-0002-0764-9606},
Jonas Nüßlein\authorrefmark{2},
Corey O'Meara\authorrefmark{3}\orcidlink{0000-0001-7056-7545},
Ivan Angelov\authorrefmark{4},
Benedikt Wimmer\authorrefmark{1}\orcidlink{0009-0004-5481-594X},
Kumar Ghosh\authorrefmark{3}\orcidlink{0000-0002-4628-6951},
Giorgio Cortiana\authorrefmark{3},
and Claudia Linnhoff-Popien\authorrefmark{2}
}}
\address[1]{Aqarios GmbH,
Prinzregentenstraße 120, 81677 Munich, Germany (email: david.bucher@aqarios.com)}
\address[2]{Mobile and Distributed Systems Chair, Ludwig-Maximilians University, Munich, Germany (email: jonas.nuesslein@ifi.lmu.de)}
\address[3]{E.ON Digital Technology GmbH, Tresckowstraße 5, Hannover, Germany (email: corey.o'meara@eon.com)}
\address[4]{Comsysto Reply GmbH, Riesstraße 22, 80992 Munich, Germany (email: i.angelov@reply.de)}
\tfootnote{This work was supported by the German Federal Ministry of Education and Research under the funding
program "Förderprogramm Quantentechnologien -- von den Grundlagen zum Markt" (funding program quantum technologies---from basic research to market), project Q-Grid, 13N16177.}


\corresp{Corresponding author: David Bucher (email: david.bucher@aqarios.com).}

\begin{abstract}
Demand Side Response (DSR) is a strategy that enables consumers to actively participate in managing electricity demand. It aims to alleviate strain on the grid during high demand and promote a more balanced and efficient use of (renewable) electricity resources. We implement DSR through discount scheduling, which involves offering discrete price incentives to consumers to adjust their electricity consumption patterns to times when their local energy mix consists of more renewable energy.
Since we tailor the discounts to individual customers' consumption, the Discount Scheduling Problem (DSP) becomes a large combinatorial optimization task.
Consequently, we adopt a hybrid quantum computing approach, using D-Wave's Leap Hybrid Cloud. 
We benchmark Leap against Gurobi, a classical Mixed Integer optimizer in terms of solution quality at fixed runtime and fairness in terms of discount allocation.
Furthermore, we propose a large-scale decomposition algorithm/heuristic for the DSP, applied with either quantum or classical computers running the subroutines, which significantly reduces the problem size while maintaining solution quality. Using synthetic data generated from real-world data, we observe that the classical decomposition method obtains the best overall \newp{solution quality for problem sizes up to 3200 consumers, however, the hybrid quantum approach provides more evenly distributed discounts across consumers.}
\end{abstract}

\begin{keywords} 
Demand Side Response, Problem Decomposition, Smart Grids, Quadratic Unconstrained Binary Optimization (QUBO), Quantum Annealing, Quantum Computing.
\end{keywords}

\titlepgskip=-15pt

\bstctlcite{IEEEexample:BSTcontrol}

\maketitle

\input{sections/00_introduction}

\input{sections/01_related_work}

\input{sections/02_problem_formulation}

\input{sections/03_decomposition}

\input{sections/05_evaluation}

\input{sections/06_conclusion}

\appendix
\input{sections/a_glossary}

\bibliographystyle{IEEEtranDoi}
\bibliography{bstcontrol,references}
\EOD

\end{document}

%% file: sections/00_introduction.tex
\section{Introduction}\label{sec:introduction}
The rising demand for energy resources and the growing adoption of renewable electricity sources have prompted a search for innovative solutions to optimize energy consumption in order to reduce grid congestion and carbon emissions. Demand Side Response (DSR)~\cite{DSR_00} has emerged as a promising strategy that focuses on actively managing and adjusting energy consumption patterns in response to grid conditions. Various studies in literature explore DSR, detailing its impact on smart grid technology \cite{DSR_01}, load scheduling~\cite{DSR_02}, energy economics~\cite{DSR_03}, as well as optimal control and pricing schemes~\cite{DSR_04}.

Price adjustment serves as a straightforward method to influence consumer behavior. With the emergence of smart devices and the electrification of heating and transportation, the response to price incentives can be progressively automated.
Typically, DSR is achieved by handing out a dynamic price to all customers simultaneously. However, the diverse usage patterns among consumers may favor alternative dynamic pricing policies. Therefore, we aim to find individual price patterns on a per-customer basis to achieve an optimal load shift. We call the distribution of discounts or penalties to specific customers the Discount Scheduling Problem (DSP). The number of customers to be considered in such a problem, i.e., an urban power grid, can become prohibitively large to be solved by classical resources.

In recent years, Quantum Computing (QC) has garnered significant attention as a potential game-changer in various domains, including optimization. Leveraging the principles of quantum mechanics, quantum optimization algorithms are hypothesized to solve complex optimization problems more efficiently than their classical counterparts.
Besides gate-based universal quantum computing, Adiabatic Quantum Computing (AQC) has emerged, which can be shown in general to be equivalent to gate-based approaches\cite{PhysRevLett.99.070502}. Quantum Annealing (QA)~\cite{RevModPhys.80.1061, RevModPhys.90.015002}, a subset of AQC, has been widely adopted for solving optimization problems~\cite{Kadowaki:1998, FINNILA1994343}. As the industry leader in quantum annealing hardware, D-Wave's quantum annealer is employed in this work to optimize the DSP. The limited size of current quantum computing hardware forces us to utilize hybrid quantum computing approaches, like Leap, which is a Cloud service offered by D-Wave and is based on internal problem size reduction~\cite{mcgeoch}. In this work, we additionally develop a customized hybrid approach that performs a problem-specific decomposition.

This paper aims to investigate the applicability of QA to DSP optimization and benchmarks the performance of hybrid quantum-classical routines against purely classical counterparts.
The overall structure is as follows: After giving a concise literature review in Sec.~\ref{sec:literature_review}, we describe the problem formulation and mathematical modeling of the DSP as a Quadratic Integer Programming (QIP) problem in Sec.~\ref{sec:problem_formulation}. Since the problem should be solvable for a customer count in realistic scenarios, Sec.~\ref{sec:decomposition} motivates and develops a problem-specific decomposition algorithm for problem size reduction. This decomposition routine proves to be very effective, as the benchmarking of classical and quantum-enhanced solvers, in Sec.~\ref{sec:experiments} shows. Finally, in Sec.~\ref{sec:conclusion} we discuss the overall summary of the work and the implications of applied quantum computing to large scale optimization problems in industry targeted to increasing renewable energy usage.

\newp[Furthermore, we observe that Gurobi, as a mixed-integer classical solver, reaches a performance limit for larger problem sizes within a fixed allocated runtime, whereas D-Wave's \texttt{Leap} Hybrid Quantum solver is able to provide results with acceptable solution quality within the allocated runtime. Nevertheless, the custom decomposition routine aided by a classical solver provides the overall best results.]{}

%% file: sections/01_related_work.tex
\section{Literature Review}\label{sec:literature_review}

\subsection{Related Work}\label{sec:related_work}

Recently, quantum computing applications in the power and energy sector~\cite{QC_PA_01, QC_PA_02, QC_PA_03, QC_PA_04} are gaining attention for the development of smart grid technology. Several important problems are addressed using quantum computing, for example power flow calculations~\cite{QC_PF01, QC_PF_02} or energy grid classification~\cite{QC_cluster}. The traditional planning and scheduling tasks in power systems, such as the minimization of generation cost or the maximization of revenue from electricity generation, are generally formulated as combinatorial optimization problems, which are often NP-hard. Using quantum-inspired optimization algorithms is expected to outperform their classical counterparts~\cite{dwave-speedup, QC_PA_02}. A wide range of optimization problems can be converted into quadratic unconstrained binary optimization (QUBO) problems~\cite{lucas2014}, which can be efficiently solved with the Quantum Approximate Optimization Algorithm (QAOA)~\cite{farhi2014a} using gate-based universal quantum computers or using D-Wave quantum annealers. In the literature, there exist multiple quantum computing approaches towards unit commitment~\cite{QC_O_02, QC_O_03, QC_O_04, QC_O_05} and other mixed integer problems~\cite{QC_O_01}, using quantum-inspired ADMM~\cite{Qc_ADMM} or Benders' decomposition methods~\cite{Qc_Benders}. Quantum annealing approaches are also used for community detection in electrical grids~\cite{QC_O_09}, peer-to-peer energy trading~\cite{QC_O_08} or coalition structure optimization~\cite{QC_coalition, QC_coalition_2}. \newp{Several research studies benchmark the performance of classical algorithms vs. hybrid quantum-classical algorithms such as Leap on large-scale instances. These include transport robot scheduling~\cite{leib2023}, job shop
scheduling ~\cite{geitz2022}, power network partition~\cite{colucci2023} and SAT problems~\cite{oshiyama2022}.}

As one of this work's main contributions is developing a problem-specific decomposition method to solve large instances of the DSP on currently available hardware, we give a brief overview of combinatorial problem decomposition algorithms in the context of quantum optimization here.
Divide-and-conquer approaches have been used for various problem instances, such as the MaxClique problem~\cite{chapuis2017, hahn2017, pelofske2019, pelofske2020}, Minimum Vertex Cover~\cite{pelofske2019a, pelofske2020}, Community Detection~\cite{guerreschi2021} and MaxCut~\cite{guerreschi2021, zhou2023}. They all combine the splitting of the problem into sub-problems using problem-related methods. In special cases, such as Ref.~\cite{zhou2023}, quantum optimization is further utilized in recombining the solution because of the special $\mathbb{Z}_2$ symmetry of MaxCut solutions. Quantum Local Search (QLS)~\cite{tomesh2022a} takes local sub-problems of a graph-based problem and iteratively improves a global solution vector. Although applicable to any graph-based problem, QLS has been specifically tested for the Maximum Independent Set problem.
The recent emergence of distributed quantum computing has led to the development of decomposition algorithms that still allow for a limited amount of quantum information exchange between the optimization of the sub-problems~\cite{saleem2022, fujii2022a}, which was successfully demonstrated for the Maximum Independent Set problem. Apart from problem-specific methods, general QUBO decomposition methods have been devised, like QBSolv~\cite{booth2017a}. Here, subsets of binary variables of the full QUBO are selected as sub-problems, which are solved on a quantum annealer, while in parallel, a classical heuristic optimizes the original problem. During the process, solutions to the sub-problems will incrementally improve the current solution state of the heuristic.


\subsection{Introduction to Quantum Annealing}\label{sec:quantum_annealing}
Quantum annealing (QA) is a heuristic for solving combinatorial
optimization problems, first proposed in 1998 by Kadowaki and Nishimori~\cite{Kadowaki:1998}.
QA utilizes the adiabatic theorem to find the unknown ground state of an Ising Hamiltonian $\mathcal{H}_\text{Ising}$, whose minimal energy state corresponds to the solution of a target problem.

With $\mathcal{H}_\text{Init}$ being the initial Hamiltonian, the annealing process can be described by the following dynamic Hamiltonian:
\begin{gather}
\mathcal{H}(s) = A(s)\mathcal{H}_\text{Init} + B(s) \mathcal{H}_\text{Ising}\\ \quad
\mathcal{H}_\text{Init} = - \sum_{i} {\sigma}_x^{i} \\ \quad \mathcal{H}_\text{Ising} = - \sum_{i} h_i {\sigma}_{z}^{i} - \sum_{i>j} J_{ij} {\sigma}_z^{i} {\sigma}_z^{j},
\end{gather}
where ${\sigma}_{x,z}^{(i)}$ are Pauli matrices operating on qubit $i$, and $h_i$ and $J_{i,j}$ are the qubit biases and coupling strengths, which encode the specific problem. $A(s)$ and $B(s)$ are known as the annealing schedule, with $s \in [0, 1]$. At $s=0$, $A(s) \gg B(s)$, while  $A(s) \ll B(s)$ for $s=1$. As we increase $s$ from 0 to 1, the system undergoes a gradual change from $\mathcal{H}_\text{Init}$ to $\mathcal{H}_\text{Ising}$. The adiabatic theorem of quantum mechanics states that if that evolution happens slowly enough and the system is initialized in the trivial ground state of $\mathcal{H}_\text{init}$, then the state will remain in the ground state of the momentary Hamiltonian~\cite{born1928}. Eventually, at $s = 1$, the state will be in the ground state of the $\mathcal{H}_\text{Ising}$. Finding the ground state of the Ising model is isomorphic to QUBO~\cite{lucas2014}, therefore, measuring the final state will reveal the solution to an NP-hard optimization task.

In quantum annealing, this transition speed will typically be faster than required for the adiabatic theorem, due to practical considerations. Nevertheless, experimental evidence suggests that, depending on the spin glass model, faster evolution times still output the optimal solution with high probability~\cite{johnson2011}. Thus, measuring the output repeatedly will eventually find the correct solution.

%% file: sections/02_problem_formulation.tex
%
%
%
%
%
%
%
\section{Discount Scheduling Problem Formulation}\label{sec:problem_formulation}
\newp{
\begin{table}[t]
    \centering
    \begin{tabular}{cc}
        Normalization Constant & Expression\\
        \hline
         $\mathcal{N}_0$ & $E(0) - E_\text{min}$  \\
         $\mathcal{N}_1$ & $N_C \zm^2 $  \\
         $\mathcal{N}_2$ & $4 N_C (N_T - 1) \zm^2  $  \\
         $\mathcal{N}_3$ & $N_C N_T \zm^2 $
    \end{tabular}
    \caption{Normalization constants for the penalty terms.}
    \label{tab:normalizations}
\end{table}

Given a discrete time horizon of $N_T$ steps $t$, a set of $N_C$ customers $c$ with projected consumption data $d_{c,t}$, and the local \coo\ grid intensity $I_t$ $[\text{g}/\text{kWh}]$, the goal of the DSP is to assign each customer individual discrete discounts $z_{c,t} \in Z$, such that the overall \coo\ emissions are minimized, but the overall consumption remains equal. Furthermore, grid constraints must be satisfied at any timestep, and the overall consumption deviations of a single customer should be kept to a minimum.
The discount categories $Z$ are defined as a symmetric set $\max Z = -\min Z = \zm$, where positive discounts are referred to as penalties, e.g. $Z = \{-0.5, -0.25, 0, 0.25, 0.5\}$.

Reducing \coo\ emissions has the advantage of increasing consumption during periods of abundant local renewable energy production. However, any other linear or quadratic function constructed from the discounts can be used as an objective for the DSP (e.g. minimizing the operational costs based on spot market prices).
}

\subsection{Preliminary Considerations}\label{sec:preliminary_considerations}

Since the distribution system operator (DSO) cannot yet automatically influence the consumption of the customer at a certain time, we have to go the detour over price incentives. We assume customers are strictly economically motivated, i.e., they alter their consumption based on price. Of course, the convenience of having access to electricity at all times is more important than saving on the cost, such that, in reality, customers cannot vary their consumption arbitrarily at any given time. However, with the emergence of electric vehicles (EVs) with home charging and heat pumps, automatically varying the load becomes possible. The given discounts then act as a protocol that communicates to a smart home appliance on the customer side when to use electricity and when not, e.g., start or stop charging the EV.


\newp{The central assumption of the DSP is that a given discount (or penalty) influences the customers to increase (or decrease) their consumption proportionally.} The consumption changes as follows
\begin{align}\label{eq:new_consumption}
    \tilde{d}_{c,t} = (1 - \chi_c z_{c,t}) d_{c,t},
\end{align}
when given a discount $z_{c,t}$. The constant $\chi_c$ is the (negative) price elasticity of demand of customer $c$. I.e., the higher $\chi_c$ is, the more customer $c$ responds to price incentives (lower its demand if price increases and vice versa). In principle, the price elasticity takes positive values below one, where $\chi_c = 1$ means a full reflection of the price change on the load change. In literature, different estimations of the electricity demand price elasticity have been made, reaching values between $0.65$ and $0.85$ in residential U.S. customers~\cite{alberini2011} and $0.8$ to $0.9$ in Swiss households~\cite{filippini2011}. A metastudy~\cite{dutta2017} on dynamic pricing reveals that the short-term price elasticity has to be estimated lower than the long-term elasticity. Nevertheless, response to dynamic pricing may be increased by automation of the load of smart devices and other enabling technologies~\cite{dutta2017}. In reality, price elasticity will vary between customers, so we formulate it as a customer-specific value. The DSO can measure the response of individual customers and adjust the elasticities for a more accurate model.

\newp{
Discrete discounts allow users to change their behavior more distinctly. For instance, providing a small discount to a thousand customers might not necessarily have the intended effect, then supplying only a few customers with moderate discounts can have. Therefore, restricting to a discrete set of categories $Z$ is sound.
}

\subsection{Mathematical Formulation}\label{sec:constriants}

\newp{

Collecting the considerations, we can finally formulate the DSP optimization problem as QIP for minimizing \coo\ production through load shifting. All terms and constraints will be explained separately in the following sub-sections.
\begin{align}
    \text{minimize:}\quad&\nonumber\\
    C(z) = &\frac{1}{\mathcal{N}_0} \sum_{c,t} I_t  (1 - \chi_c z_{c,t}) d_{c,t} \label{eq:objective}  \\
    + &\frac{\lambda_1}{\mathcal{N}_1}\sum_{c}  \left(\frac{1}{D_c}\sum_t d_{c,t} z_{c,t}\right)^2\label{eq:sconstr_consumption_deviation}\\
    + &\frac{\lambda_2}{\mathcal{N}_2} \sum_{c,t} (z_{c,t} - z_{c,t + 1})^2 \label{eq:discount_change_penalty}\\
    + &\frac{\lambda_3}{\mathcal{N}_3}\sum_{c,t} z_{c,t}^2 \label{eq:discount_regularization}\\
    \text{such that:}\quad& \nonumber\\
    \sum_{c,t}& z_{c,t} d_{c,t} = 0,\label{eq:constr_global_zero}\\
     -\Delta p_t \leq \sum_{c}& \chi_c z_{c,t} d_{c,t} \leq \Delta p_t \quad \forall t \in \{1,\dots, N_T\}.\label{eq:constr_power_restriction}
\end{align}
Here, $C(z)$ is the cost function to be minimized, and $D_c$ is the total power draw of a customer $D_c = \sum_t d_{c,t}$. Furthermore, $\lambda_i$ refers to penalty factors and $\mathcal{N}_i$ to normalization constants employed to keep the impact of the penalty factors independent of problem size and data. The normalization constants are chosen such that the effect of each penalty term is 1 if all discounts are assigned in the worst possible way; see Table~\ref{tab:normalizations}. 

The formulation of the objective and the purpose of all penalty terms and constraints present in the problem statement~\eqref{eq:objective}--\eqref{eq:constr_power_restriction} will be explained in the following sub-sections.

\subsubsection{\coo\ Emission Minimization}
The combined \coo\ emission is proportional to the changed consumption~\eqref{eq:new_consumption}
\begin{align}\label{eq:co2_reduction}
    E(z) = \sum_{c,t} I_t [1 - \chi_c z_{c,t}]d_{c,t},
\end{align}
and serves as the main objective of the minimization formulation. The normalization constant $\mathcal{N}_0$ is chosen to map the range of \coo\ emissions between 0 and 1. Therefore, we utilize the trivial origin configuration $P(z = 0)$ as the maximal value and compute a naive lower bound for the \coo\ emissions to set $\mathcal{N}_0 = E(0) - E_\text{min}$:
\begin{align}
    E_\text{min} = \sum_{c,t} \left[1 - \chi_c \sign(I_t - \langle I_t \rangle) z_\text{max} \right] d_{c,t},
\end{align}
which gives all customers the full discount if $I_t$ is smaller than the average and the full penalty if $I_t$ is larger, respectively.

Note that this lower-bound solution does not satisfy the constraints of the formulation~\eqref{eq:constr_global_zero},~\eqref{eq:constr_power_restriction}. Therefore, it is substantially smaller than the actual best solution.


\subsubsection{Consumption deviation penalty}

Customers should not change the total energy they consume over the optimization horizon, i.e.,
\begin{align}
    \sum_{t} d_{c,t} z_{c,t} \approx 0 \quad \forall c \in \{1,\dots\,N_C\}.
\end{align}
A perfect equality can generally not be achieved because of the discrete discounts, except for the trivial case $z_{c,t} = 0$. Therefore, it is represented by the penalty term Eq.~\eqref{eq:sconstr_consumption_deviation} as a quadratic soft-constraint with penalty factor $\lambda_1$.

\subsubsection{Discount change penalty}
As discussed in Sec.~\ref{sec:preliminary_considerations}, longer periods with similar discounts exhibit better customer response. We, therefore, employ a penalty function that tries to minimize consecutive discount changes in Eq.~\eqref{eq:discount_change_penalty}. The corresponding penalty factor $\lambda_2$ will be chosen small ($\lambda_2 < \lambda_1$).

\subsubsection{Discount regularization} 
We attempt to assign tarif discounts that affect the objective function $C$ by a large enough amount. Suppose a customer consumes an equal amount at two timesteps with $I_{t_1} = I_{t_2}$. Assigning $z_{c,t_1} = -z_{c,t_2} = \zm$ would not change the cost compared to $z_{c,t_1} = z_{c,t_2} = 0$, but can be given anyways. A small $L2$-regularization, see Eq.~\eqref{eq:discount_regularization}, ensures that discounts are only given if they benefit the overall goal, with $\lambda_3 \leq \lambda_2$. $L2$-regularization is chosen over $L1$-regularization since it naturally maps into QUBO.

\subsubsection{Global consumption deviation constraint}
Even though the per-customer consumption deviation is soft-constrained~\eqref{eq:sconstr_consumption_deviation}, the consumption deviation of all customers together can be zero up to numeric precision. Globally, i.e., the combined view of all customers, we do not want any change in overall consumption. Hence, it is a hard constraint, see Eq.~\eqref{eq:constr_global_zero}.

\subsubsection{Power restriction constraint}
The momentary change in consumption (power restriction constraint) of all customers combined must be bounded due to grid voltage peaks and therefore the hard-constrained Eq.~\eqref{eq:constr_power_restriction} has been introduced. Additionally, for load shifting, we require a time-window where consumption can be increased and decreased. The values for $\Delta p_t$ can be determined using power flow computations and can, in principle, also be asymmetric. Of course, the presented power restriction is a simplification, but it suffices for an initial investigation of the problem.

\subsection{Discount Encoding}

Discrete discounts $z_{c,t} \in Z$ offer another benefit, which is that they can relatively easily be encoded through binary variables~\cite{lucas2014}. This makes translating the problem formulation into QUBO easier, which is required for employing quantum optimization techniques.

We will focus on \emph{integer encoding} of the discount set $Z$: Here, we discretize the range $[-z_\text{max}, z_\text{max}]$ into $N_K$ linearly spaced categories. Generally, the range can also be asymmetric but is not considered in this work. 
Therefore, $Z = \{-\zm + i \Delta z\,|\, i = 0,\dotsc\,N_K-1\}$, with $\Delta z = \frac{2 \zm}{N_K-1}$. This range can subsequently be expressed using $Q = \lfloor \log_2 N_K + 1\rfloor$ binary variables $x_{c,t,k}$ for each discount~$z_{c,t}$
\begin{gather}
    z_{c,t} = \Delta{z} \sum_{k=0}^{Q-1} w_k x_{c,t,k} - \zm, \\
    \text{with }w_k = \begin{cases}2^k &\text{if } k < Q - 1 \\ N_K - 2^{Q-1} + 1  &\text{else.} \end{cases}
\end{gather}
Every bit combination $x_{c,t,k}$ results in a valid encoding, making an additional penalty term for encoding obsolete~\cite{lucas2014}. This encoding is very space efficient, allowing an exponential number of categories to be represented with a linearly growing number of qubits. 

An alternative method for encoding discounts would be \emph{one-hot encoding}, where $N_K$ bits encode every item of $Z$ by only setting one bit to 1 and the other ones to zero. This is a more general framework that allows any $Z$ (not just linearly spaced) to be encoded through binary variables.
However, it requires $N_k$ binary variables per customer and an additional constraint. In the context of QUBO, that constraint has to be enforced as an additional penalty term. Preliminary experiments have shown advantageous results for integer encoding compared to one-hot encoding.

\subsection{On customer savings} \label{sec:on_customer_savings}

Given the customers initially receive a flat tariff $v_0$ $[\text{€}/\text{kWh}]$, the discount or penalty ($v_0 \rightarrow (1 + z_{c,t}) v_0$) only affects the consumption that deviates from the projected consumption $\tilde{d}_{c,t} - d_{c,t}$. Consequently, the customer pays
\begin{align}\label{eq:price_change_general}
v_{c,t} \tilde{d}_{c,t} = v_0 d_{c,t} +  (1 + z_{c,t}) v_0 (\tilde d_{c,t} - d_{c,t}).
\end{align}
for a specific timestep.

The customer's cost change over the optimization horizon can be computed via the sum of momentary price differences through  Eq.~\eqref{eq:price_change_general}:
\begin{align}
\Delta v_{c} = \sum_t (v_0 - v_{c,t}) \tilde{d}_{c,t} = v_0 \sum_t z_{c,t} (\tilde d_{c,t} - d_{c,t}).
\end{align}
Note that we have used the sum over the changed consumption as the baseline for our comparison, since in any case $\sum_t d_{c,t} \approx \sum_t \tilde{d}_{c,t}$ and we only want to compare the cost for the same amount of purchased electricity.

Substituting in the altered consumption from Eq.~(\ref{eq:new_consumption}), we obtain a change in cost given by
\begin{align} \label{eq:price_change}
    \Delta v_c = -v_0\chi_c \sum_t z^2_{c,t}d_{c,t}.
\end{align}
The absolute price change is dependent on the flat tariff and the total consumption of the customer. We will therefore look at the relative savings $s_c = -\Delta v_c / \sum_t v_0 \tilde{d}_{c,t} \geq 0$ in the experiments section.

As $z_{c,t}^2 \geq 0$ and $\chi_c \geq 0$, the customer's price change is guaranteed to be $\Delta v_c \leq 0$ negative, so a customer will always save money by complying with the incentives. The savings are exactly zero if the customer does not respond to incentives at all, i.e., $\chi_c = 0$. 

\subsection{Grid Data}

For the DSP, we require forecasted consumption data $d_{c,t} \geq 0\,[\text{kWh}]$ for each customer and predicted grid \coo\ intensity $I_t$ of the power generation in the considered region. We use standard load profiles of residential and industrial customers, which we modify by adding noise and shifting in time. Additionally, the load profiles get scaled to resemble various numbers of residents. Moreover, we include photovoltaic (PV) electricity generation by estimating the potential based on roof data of Munich residential areas and simulating the production from historical solar irradiance data. PV production reduces the customers' consumption. Grid infeed, i.e., if more PV is generated than consumed, is not specially considered. The grid \coo\ intensity is taken from the real-world data in Munich.%
\footnote{The data is provided by E.ON's App for monitoring local \coo\ intensities: \url{https://www.bayernwerk.de/de/fuer-zuhause/oekoheld.html}}.
The data used throughout this text consists of roughly 16000 customers and 76 timesteps, spanning a 19-hour period with 15-minute intervals. The \coo\ and solar data are from January 13, 2023.
}

%% file: sections/03_decomposition.tex
%
%
%
%
%
%
%
%
%
\section{Problem Decomposition}\label{sec:decomposition}

The number of integer variables needed to construct the discount matrix is $N_C \times N_T$. Given a one-day optimization horizon with 15-minute timesteps, each customer requires 96 integer decision variables in the problem. However, as the number of customers will grow quite large%
\footnote{Typically, we want to consider more than 1000 customers.},
the number of integers grows akin. Even worse,
the number of qubits in the quantum formulation is scarce, and every integer must be encoded with $Q$ qubits. Thus, the move to a hybrid quantum-classical optimization scheme seems inevitable.

In this section, we propose a hybrid approach that is based on problem decomposition. Despite the drawback that decomposition increases solution bias, we find that we can manage the hard constraints of the DSP classically in a pre-processing step. This eliminates the need for a costly reformulation of inequality constraints with slack variables.
Fig.~\ref{fig:decomposition_overview} shows an overview of the steps taken for the decomposition.

\subsection{Motivation}\label{sec:motiviation}

\subsubsection{Global Solution}\label{sec:global_solution}
Shifting the perspective from the individual customer level to a global scope, where all customers are regarded as a unified entity,
we consider the overall consumption $D_t = \sum_{c}d_{c,t}$ and the mutable consumption, i.e., the consumption weighted by the individual customer susceptibilities $\widetilde{D}_t = \sum_c \chi_c d_{c,t}$. Furthermore, we can express the weighted average of all discounts given per customer---from now on called \emph{effective discount}---as follows
\begin{align}\label{eq:effective_discount}
    \zeta_t = \langle z_{c,t} \rangle_c = \frac{1}{\widetilde D_t} \sum_c \chi_c d_{c,t} z_{c,t} \in [-\zm, \zm].
\end{align}
Utilizing the formulation of the effective discount, we can transform the \coo\ production from Eq.~(\ref{eq:co2_reduction}) into
\begin{align}
    E(\zeta) = \sum_t I_t \left( D_t - \widetilde D_t \zeta_t \right).
\end{align}

The global consumption deviation constraint, Eq.~\eqref{eq:constr_global_zero}, and the power restriction constraint, Eq.~\eqref{eq:constr_power_restriction} can be expressed solely in terms of the effective discount. Therefore, we represent the global version of the DSP as a linear program
\begin{align}\label{eq:global_dsp}
\begin{split}
\text{minimize:} \quad &E(\zeta) \\
\text{such that:} \quad  &-\Delta p_t \leq \widetilde{D}_t \zeta_t \leq \Delta p_t \quad \forall t \in {1,\dots,N_T} \\
                & \sum_t \widetilde{D}_t \zeta_t = 0.
\end{split}
\end{align}
This formulation disregards any per-customer constraints that are still part of the DSP. Nevertheless, it is a useful tool to estimate how much \coo\ reduction is maximally possible with all the hard constraints \eqref{eq:constr_global_zero}, \eqref{eq:constr_power_restriction} in place.
In fact, the solution $\zeta_t^*$ is guaranteed to give an optimal lower bound $E(\zeta^*)$ 
\begin{align}\label{eq:global_solution}
    E(\zeta^*) \leq E(z) \quad \forall z \in \mathcal{Z},
\end{align}
where $\mathcal{Z} = \{z \in Z^{N_C \times N_K} \text{ s.t. } \eqref{eq:constr_global_zero}, \eqref{eq:constr_power_restriction} \text{ hold}\}$ is the set of feasible discount matrix configurations. The global DSP consists of only $N_T$ continuous variables. Thus, it can be quickly and efficiently solved using standard procedures like the Simplex method~\cite{nelder1965}.

Given an optimal effective discount, $\zeta_t^*$, we can utilize Eq.~(\ref{eq:effective_discount}) to optimize the integers $z_{c,t}$ for the individual customers per timestep, i.e. $\min_z [\zeta_t(z) - \zeta_t^*]^2$. Additionally, we can include the penalty terms from the DSP \eqref{eq:sconstr_consumption_deviation}--\eqref{eq:discount_regularization} in the subsequent optimization. However, doing so would yield an optimization problem the same size as the original problem.

Nonetheless, the following section reveals that we can achieve a satisfactory approximation of a continuous effective discount by considering only a limited number of customers. As a result, we can divide the customers into smaller groups or chunks and optimize each chunk separately.

\subsubsection{Representational Power}

\Figure()[width=3.3in]{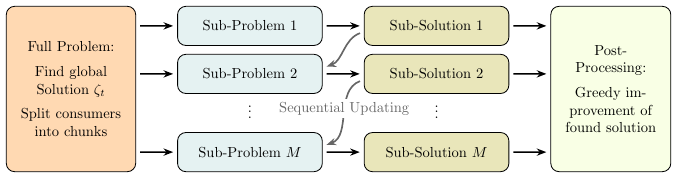}
    {Overview of the decomposition routine. The problem is split into sub-problems. Solutions can influence the following sub-problems via sequential updating. Finally, sub-solutions are gathered to a full solution and a post-processing step is employed that improves the solution quality greedily while also making the power restriction constraint is satisfied.\label{fig:decomposition_overview}}

\begin{figure}
\centering
\includegraphics[width=3in]{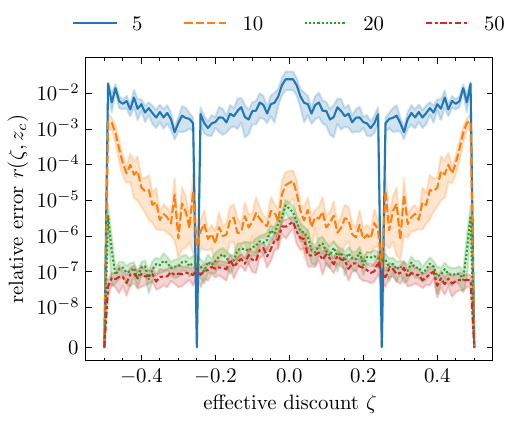}%
\caption{The relative approximation error for different values of $\zeta$ averaged over multiple timesteps. The different colors show the approximation error for an increasing number of customers averaged over 20 timesteps.
The central peaks are due to absolute errors getting amplified at small effective discounts. Because the effective discounts $\zeta = \pm 0.25, \pm 0.5$ are exactly realizable by giving all customers $\pm 25\%, \pm 50\%$ discounts, we can observe spikes there. Even though zero error can be achieved in the $\zeta = \pm 0.25$ case, Gurobi finds different configurations with good enough error first (for larger than five customer chunks). Effective discount $\zeta = 0$ is not shown since the relative error is not defined.} 
\label{fig:global_soultion_accuracy}
\end{figure}

In this section, we motivate that Eq.~\eqref{eq:effective_discount} can be fulfilled for any arbitrary $\zeta_t$ with sufficient accuracy given a small constant number of customers.
We will focus on a discount range $\zeta_t \in [-1/2, 1/2]$ and five discrete discounts $z_{c} \in \{-1/2, -1/4, 0, 1/4, 1/2\}$.
From the generated consumption data, see Sec.~\ref{sec:experiments}, we take a random set of customers and compute
\begin{align}
    \min_{z_{c}} r(\zeta, z_c) = \min_{z_c} \frac{1}{\zeta} \left| \frac{1}{D} \sum_{c} d_{c,t} z_{c} - \zeta\right|
\end{align}
for all available timesteps. 
Fig.~\ref{fig:global_soultion_accuracy} shows the result with different numbers of customers. The average over all timesteps is plotted, and the error bands indicate a 95\% confidence interval.
It is evident that even with only ten customers, the relative error remains consistently below 1\%. As more customers are added, the error decreases significantly, reaching a negligible level. 
Therefore, we contend that by maintaining a small, constant number of customers within a chunk (e.g., 20-50 customers), it is possible to obtain a reliable approximation of an effective discount while still considering the per-customer soft-constraints of the DSP.

\subsection{The Full Decomposition Routine}

\newp{We} now assemble the pieces into a full hybrid routine for decomposition, as seen in Fig.~\ref{fig:decomposition_overview}. The process begins with solving the global DSP~\eqref{eq:global_dsp}, followed by dividing customers into chunks. We sort the customers by total consumption and split them into $M$-sized groups, s.t. the largest customers are arranged in the first chunk, etc. We argue that it is better to have customers with comparable consumption in one chunk because they can counteract each other better than e.g. one industrial customer and 20 single households.
For each chunk, we can define sub-problems in which special effective discounts per chunk are introduced in Sec.~\ref{sec:chunk_problems}. \newp{These sub-problems are of QUBO form and aim to assign discounts to customers such that the overall effect matches an effective discount while making sure, each customer does not deviate from the its total consumption by much.}

Since we can solve the sub-problems sequentially, we can enhance the results by incorporating the errors from prior optimizations into the subsequent sub-problems in Sec.~\ref{sec:sequential_updating}. Eventually, all the chunks are collected, and a final post-processing step shown in Sec.~\ref{sec:post_processing} is applied to ensure that no constraints are violated.

\subsubsection{Chunk Problems}\label{sec:chunk_problems}
The customers are partitioned into $M = N_{c} / m$ mutually exclusive chunks $C_j$, s.t. $\bigcup_j C_{j} = \{1, \dots, N_C\}$ and $C_i \cap C_j = \emptyset\, \forall i \neq j$. Note, that we require and expect the chunk size to be chosen, s.t. $N_C \mod m = 0$.

Most likely the consumption deviation per chunk
\begin{align}
    \sum_{c \in C_j} \sum_t \chi_c d_{c,t} \zeta^*_t \neq 0 \quad \forall j
\end{align}
is not zero, which, by default, introduces a bias in the consumption deviation soft-constraint~\eqref{eq:sconstr_consumption_deviation}. Thus, the first goal is to define chunk effective discounts $\xi_t^j$ with the following properties:
\begin{align}
    \sum_t \widetilde{D}_t^j \xi_t^j &= 0 \quad \forall j, \label{eq:ced_cond1}\\
    \frac{1}{\widetilde{D}_t} \sum_{j=1}^M\widetilde{D}^j_t \xi_t^j &= \zeta^*_t \quad \forall t\label{eq:ced_cond2},
\end{align}
where we define an alterable consumption for one chunk $\widetilde{D}^j_t = \sum_{c \in C_j} \chi_c d_{c,t}$, similar to the definition of the total mutable consumption. 

We define the chunk-effective discount as follows
\begin{align}
    \xi_t^j = \zeta^*_t - \frac{\alpha_t}{\widetilde D_t^j} \sum_{t'} \widetilde D_{t'}^j \zeta_{t'},
\end{align}
where $\alpha_t$ are arbitrarily chosen constants, s.t. $\sum_{t} \alpha_t = 1$. The conditions \eqref{eq:ced_cond1} and \eqref{eq:ced_cond2} are satisfied with this definition. The values $\alpha_t$ are chosen constant $\alpha_t = 1/N_T$, but we have to make sure that $\xi_t^j \in [-\zm, \zm]\, \forall t, j$. If this is not possible for one timestep $t$, $\alpha_t$ has to be reduced while the remaining $\alpha$s have to be increased.

The optimization objective is to approximate the following equality with the chunk effective discount as
\begin{align}
    \widetilde D^j_t \xi^j_t = \sum_{c \in C_j} d_{c,t} \chi_c z_{c,t}\quad\forall t \in \{1, \dots, N_T\}.
\end{align}
The objective can be reformulated as a least squares error problem to find an optimal $z^*_{c,t}$ 
\begin{align}\label{eq:sub_problem}
   \argmin_{z_{c,t}} &\frac{1}{N_T \zm^2}\sum_t \left(\xi_t - \frac{1}{\widetilde D^j_t} \sum_{c \in C_j} d_{c,t} \chi_c z_{c,t} \right)^2
\end{align}
and is directly in QUBO form after the binary representation of the discounts has been substituted into the formulation.

The previously discussed penalty terms and regularizations---consumption deviation \eqref{eq:sconstr_consumption_deviation}, discount change penalty \eqref{eq:discount_change_penalty} and discount regularization \eqref{eq:discount_regularization}---can be carried over to this optimization problem.

\subsubsection{Sequential updating}\label{sec:sequential_updating}

When the sub-problems are solved in sequence, the error between the true achieved effective discount and the demanded one can be carried over into the next optimization to be corrected. For optimizing $\xi_t^j$, the procedure can be adapted as
\begin{align}
  \xi^j_t \leftarrow \xi^j_t + \frac{1}{\widetilde D_t^j}\sum_{i = 1}^{j - 1}\left(\widetilde D_t^{i} \xi^{i}_{t} - \sum_{c \in C_{i}} z^*_{c,t} d_{c,t}\right).
\end{align}
Doing so will significantly improve the overall accuracy of the method.
Of course, one has to ensure that the altered $\xi$s do not exceed the bounds $[-\zm, \zm]$.

\subsubsection{Post-processing}\label{sec:post_processing}
Finally, we describe a post-processing scheme that refines the result and ensures that the power restriction constraint~\eqref{eq:constr_power_restriction} is held.
Algorithm~\ref{alg:post} describes the greedy improvement of the solution.

\RestyleAlgo{ruled}
\begin{algorithm}[ht]
\small
\caption{The post-processing algorithm}\label{alg:post}
\SetKwComment{Comment}{\# }{}
\SetKw{Continue}{continue}
\KwData{$d_{c,t}, r \in \mathbb{N}\,\,\,$\\\Comment{$r$ is a parameter that dials the accuracy/runtime}}
\KwResult{$z_{c,t}$}
$\Delta z \gets 2\, \zm / (N_K - 1)$\Comment*{Discount step}
$\Delta_{c,t} \gets \chi_c d_{c,t} \Delta z$\Comment*{Possible deviations}
$\delta_c \gets \sum_t \chi_c d_{c,t} z_{c,t}$\Comment*{Consumption deviation}
\For{$t \in \{1, \dots, N_T\}$}{
    $p \gets \sum \chi_c d_{c,t} z_{c,t}$\Comment*{Power deviation}
    $\varepsilon \gets \zeta_t^* \widetilde{D}_t - p$\Comment*{Error from demanded}
    \Comment{Increasible customers}
    $C^{\uparrow} \gets \{ c = 1,\dots,N_C \,|\, z_{c,t} < \zm, \delta_c < -\Delta_{c,t} / 2 \}$\; 
    \Comment{Decreasible customers}
    $C^{\downarrow} \gets \{ c = 1,\dots,N_C \,|\, z_{c,t} > -\zm, \delta_c > \Delta_{c,t} / 2 \}$\; 
    $C^{\uparrow} \gets \text{limit}(C^\uparrow, r)$\Comment*{reduce size $|C^\uparrow| = r$}
    $C^{\downarrow} \gets \text{limit}(C^\downarrow, r)$\;
    \Comment{Compute combinations of increasing and decreasing two customer discounts}
    $X_{c,c'} \gets \sign(\zeta^*_t) (\varepsilon - [\Delta_{c,t} - \Delta_{c',t}])$\;
    \Comment{Find positive (feasible) ones}
    $C_2 \gets \{(c^\uparrow, c^\downarrow) \in C^\uparrow \times C^\downarrow \,|\, X_{c^\uparrow, c^\downarrow} > 0\}$\;
    \Comment{Get the best move}
    $c^\uparrow, c^\downarrow \gets \argmin_{c^\uparrow, c^\downarrow \in C_2} X_{c^\uparrow, c^\downarrow} $\;
    \If{$X_{c^\uparrow, c^\downarrow} > \sign(\zeta^*_t)\varepsilon$}{
        \Continue{}\Comment*{No improvement}
    }
    \Comment{Update solution and consumption deviation}
    $z_{c^\uparrow,t} \gets z_{c^\uparrow,t} + \Delta z$\;
    $z_{c^\downarrow,t} \gets z_{c^\downarrow,t} - \Delta z$\;
    $\delta_{c^\uparrow} \gets \delta_{c^\uparrow} + \Delta_{c^\uparrow, t}$\;
    $\delta_{c^\downarrow} \gets \delta_{c^\downarrow} - \Delta_{c^\downarrow, t}$\;
}
\end{algorithm}

Conceptually, it is quite simple: For each timestep, we extract those customers whose discounts can be increased or decreased while also improving the consumption deviation penalty~\eqref{eq:sconstr_consumption_deviation}. Then we try all combinations between one increase and one decrease and investigate how the effective discount behaves. If $\zeta^*_t$ is negative, we want the real effective discount to be as close as possible but at least larger than $\zeta^*_t$. If it is positive, the other way around. Doing so always satisfies constraint~\eqref{eq:constr_power_restriction}. We find the combination that matches the requirements the best and update the respective discounts if it achieves an improvement. Otherwise, the timestep is skipped.

Since all possible combinations of up and down moves have to be considered, the complexity of the Algorithm scales at worst with $\mathcal{O}(N_T N_C^2 / 4)$. Nevertheless, limiting the possible moves to at most $r$ provides sufficient accuracy, empirically. This then reduces the complexity to $\mathcal{O}(N_T N_C + N_T r^2)$.

%% file: sections/05_evaluation.tex

\section{Experiments \& Results}\label{sec:experiments}

\subsection{Experimental Setup}
To benchmark the performance of solving the DSP, we consider out-of-the-box solvers and our developed decomposition method and evaluate the results using a set of metrics that best represent the different goals described in the DSP formulation.

\subsubsection{Investigated Solvers}

An overview of the considered solvers and settings can be found in Table~\ref{tab:overview_of_solvers}. As a state-of-the-art purely classical baseline, we use \texttt{Gurobi}\footnote{All experiments with Gurobi were conducted on an M1 MacBook Pro (2020) with Gurobi Version 9.0}~\cite{gurobi2023}. This is compared to D-Wave’s \texttt{LeapHybridCQM} solver~\cite{mcgeoch} (called just \texttt{Leap} in the following), which is a quantum-classical hybrid algorithm that uses classical algorithms to optimize the problem while using quantum computers to solve suitable sub-tasks. This has the benefit of solving larger problems than possible directly on current quantum hardware while also supporting more sophisticated optimization models that include hard constraints. Like our decomposition routine, \texttt{Leap} partitions the problem into sub-problems via a proprietary algorithm. However, it follows a general ansatz compared to our problem-specific one. \texttt{Leap} is accessed through D-Wave's Cloud service. \newp{Both \texttt{Leap} and \texttt{Gurobi} solve the optimization problem presented in Eq.~\eqref{eq:objective}--\eqref{eq:constr_power_restriction}}. These two out-of-the-box solvers are compared against our own problem-specific decomposition routine introduced in Sec.~\ref{sec:decomposition}, subsequently called \texttt{Decomp-Gurobi}, \texttt{Decomp-Leap} or \texttt{Decomp-QPU}, depending on the method considered for solving the chunk problems~\eqref{eq:sub_problem}. \texttt{QPU} refers to direct access to the D-Wave's Quantum Annealing processor Advantage 4.1~\cite{mcgeoch}. Whenever a decomposition solver is followed by an integer, it refers to the chunk size $m$. The post-processing algorithm is used with a cut-off value $r = 500$.

In preliminary experiments, we additionally investigated D-Wave's \texttt{QBsolv} hybrid decomposition algorithm~\cite{booth2017a}, but the performance was not comparable to the approaches presented here. Furthermore, we have noticed that the solution quality did not depend on the sub-solver chosen (e.g., D-Wave's QPU or a Simulated Annealing heuristic.), indicating that the classical Tabu Search~\cite{glover1986} is solely responsible for the optimization work done.

The hybrid \texttt{Leap} solver only has the time limit as a control parameter exposed to the user. The time limit is bound from below by the minimum runtime that is heuristically calculated from the input problem (i.e., from the number of variables and couplings involved). Because it only depends on the problem structure and not the solution quality found, the minimum time limit cannot be used as the scaling metric. Furthermore, \texttt{Leap} exploits the full time limit setting and does not abort when satisfactory energy has been reached. Thus, a time-to-solution metric is not achievable within a single run and, therefore, also not considered in our benchmarking.

To alleviate the issue and ensure a fair comparison, we give each solver a heuristically increasing time limit of $0.1\,\mathrm{s} \times N_C$.
We observed that \texttt{Leap} tends to overrun the set timeout, which is the reason why we first run \texttt{Leap} with the linear growing timeout and then run the remaining solvers with the timeout matching \texttt{Leap}s runtime.
Since the decomposition solver consists of multiple sub-solver calls, we set the timeout for each sub-solver as the whole timeout divided by the number of chunks, i.e., a timeout of $0.1\,\mathrm{s} \times m$.

\begin{table}
    \centering
    \caption{Overview of the investigated Solvers}
    \label{tab:overview_of_solvers}
    \begin{tabular}[h]{c|l}
        Name & Description \\
        \hline
        \texttt{Gurobi} & Classical Solver\\
        \texttt{Leap} & Quantum classical hybrid solver\\
        \texttt{Decomp-Gurobi} & Decomposition with \texttt{Gurobi} as sub-solver\\
        \texttt{Decomp-Leap} & Decomposition with \texttt{Leap} as sub-solver\\
        \texttt{Decomp-QPU} & Decomposition with D-Waves Advantage 4.1\\&quantum annealing processor as sub-solver\\
        \texttt{Decomp-SA} & Decomposition with simulated annealing as\\&sub-solver\\
    \end{tabular}
\end{table}

\subsubsection{Metrics}

Because we consider an optimization task with multiple goals involved, it is not sufficient to consider only the objective value of our model as a performance metric. Instead, we simultaneously investigate multiple metrics:
\newp{
\begin{itemize}

\item \emph{Cost}: The cost, or objective, of the optimization problem, Eq.~\eqref{eq:objective}--\eqref{eq:discount_regularization}, is the main metric for comparing solver performance. To ensure an easier comparison between problem instances, we investigate the relative cost error with respect to the global solution from Eq.~\eqref{eq:global_solution}, defined as follows
\begin{align}
    \frac{|C(z) - C(\zeta^*)|}{|C(\zeta^*)|},
\end{align}
with $C(\zeta^*) = E(\zeta^*) / \mathcal{N}_0$. This is a guaranteed lower bound to the cost since all penalty terms are bounded from below by zero.

\item \emph{\coo\ reduction}: The \coo\ reduction is the central term in the DSP objective. Hence, it is also valuable to inspect it separately. We therefore compute the relative \coo\ reduction error through Eq.~\eqref{eq:global_solution} \begin{align}
    \frac{E(z) - E(\zeta^*)}{E(0) - E(\zeta^*)}.
\end{align}
$E(0)$ is the \coo\ emission prior to discount scheduling.

\item \emph{Consumption deviation standard deviation}: We expect the consumption deviations for each customer to be centered around zero since the problem is constrained to have a zero total consumption deviation. We therefore measure the standard deviation of the customer discount deviations as follows
\begin{align}\label{eq:consumption_deviation_metric}
    \sqrt{\frac{1}{N_C} \sum_c \left(\frac{1}{D_c} \sum_t d_{c,t} z_{c,t} \right)^2}
\end{align}

\item \emph{Average discount changes}: Since we strive to reduce the changes between two discount categories as much as possible, we measure the average discount changes:
\begin{align}
    \frac{1}{N_C (N_T - 1)} \sum_c \sum_{t = 1}^{N_T - 1} (1-\delta_{z_{c,t}, z_{c,t+1}}),
\end{align}
where $\delta$ refers to the Kronecker-Delta.

\item \emph{Average relative cost savings}: Not a quantity that is optimized for in the objective, but interesting for the DSO, is the relative cost savings per customer, as defined in Sec.~\ref{sec:on_customer_savings}. To obtain a single indicator of the performance, we evaluate the mean $\langle s_c \rangle_c$ of the relative savings.

\end{itemize}
}

\subsubsection{Parameters}
\begin{table}
    \centering
    \caption{Parameter setting for the investigated problems}
    \begin{tabular}{ccl}
         Parameter & Value & Description \\
        \hline
        $\chi_{c}$ & 1 & Customer price elasticity\\
        $\zm$ & 50\% & Maximum discount \\
        $N_K$ & 5 & Number of discounts\\
        $\Delta p_t$ & $0.1 \times \langle D_t\rangle_t $ & Power restriction \\
        $\lambda_1$ & $0.1$ & Consumption deviation penalty \\
        $\lambda_2$ & $10^{-5}$ & Discount regularization\\
        $\lambda_3$ & $10^{-4}$ & Discount change penalty
    \end{tabular}
    \label{tab:parameter_settings}
\end{table}

Solving the DSP for a given dataset, consisting of the consumption of $N_C$ customers at $N_T$ timesteps, requires fixing a set of open variables and parameters. In a real-world scenario, the customer price elasticity on demand $\chi_c$ could be measured from the individual customer's behavior. However, as it only acts as a proportionality constant, we set $\chi_c = 1$ for this investigation. Next, we use five discount categories, with a 50\% discount maximally. That, in turn refers to the following valid discounts $z_{c,t} \in \set{-50\%, -25\%, 0\%, 25\%, 50\%}$. As a consequence, a discount of, e.g., 50\% would result in an increase in the customer's consumption by 50\%.

The power deviation bounds $\Delta p_t$ are set to a constant 10\% of the average total consumption ($0.1 \times \langle D_t \rangle_t$). For the purpose of this novel problem formulation and benchmarking regarding scalability and solution quality, this is a pragmatic approach to approximation. In practice, however, those values may be derived from real-world grid constraints that can be inferred through power-flow calculations.

Finally, the remaining penalty parameters are fixed by analyzing a small-scale example with \texttt{Gurobi} and dialing in the strengths of the penalties, such that they have a reasonable effect for the \texttt{Gurobi} result. It is important to note that a comprehensive examination of the solver's response to parameter settings is beyond the scope of the current investigation.

An overview of all parameter settings is given in Table~\ref{tab:parameter_settings}.

\subsection{Example with 100 Customers}

\begin{figure}
    \centering
    \includegraphics[width=0.45\textwidth]{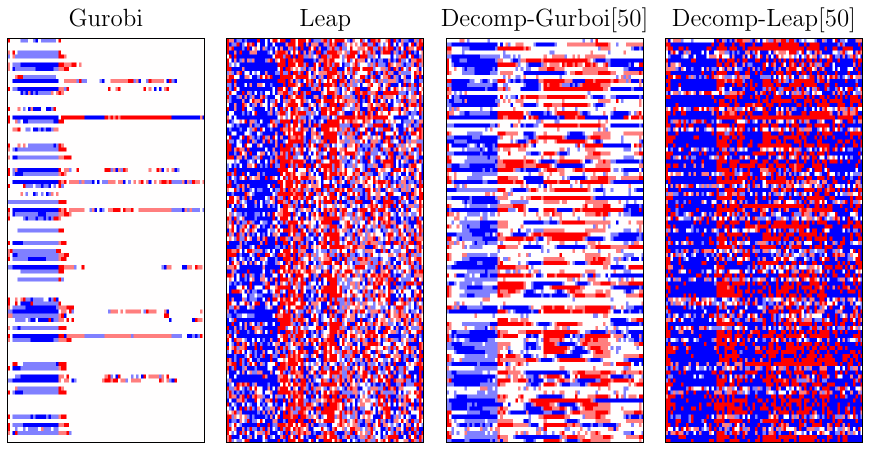}
    \caption{The discount matrices $z_{c,t}$ found by the investigated solvers for $N_C = 100$. Blue indicates a discount, and red corresponds to a penalty. White means no discount given at all. Despite their effects on the overall consumption (see Fig.~\ref{fig:ex100-consumption}) being the same, the discount matrices differ a lot from each other. It is apparent that \texttt{Gurobi} hands out the discounts more greedily than \texttt{Leap}, indicating a bigger impact of the regularization.} 
    \label{fig:ex100-discount-matrix}
\end{figure}

\begin{figure}
    \centering
    \includegraphics[width=0.4\textwidth]{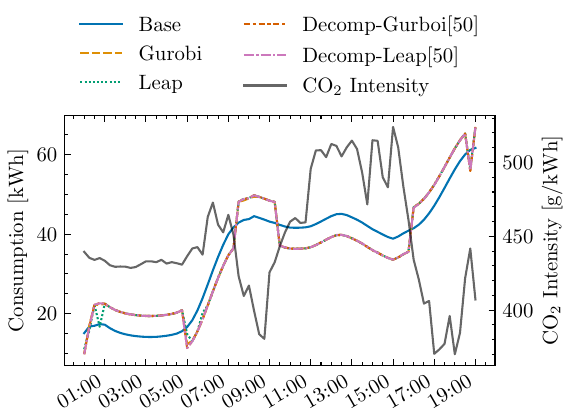}
    \caption{The effect of the DSP solution for problem size $N_C = 100$. The plot shows the aggregated consumption with and without ($z = 0$) discounts in place, as well as the grid \coo\ intensity. Visually, the solutions of all solvers produce a similar effective consumption change, as already predicted in Sec.~\ref{sec:decomposition}. As expected, times with high \coo\ emissions produce an effective decrease in consumption and vice versa.}
    \label{fig:ex100-consumption}
\end{figure}

\newp{We} first examine the optimization result of the different solvers in detail for a 100-customer example \newp{qualitatively }before focusing on the previously discussed metrics.
For that, we analyze the solutions of four solvers, \texttt{Gurobi}, \texttt{Leap}, and two $m = 50$ decomposition methods with the same solvers as the sub-routine. 
The results for the discount matrices $z_{c,t}$ can be seen in Fig.~\ref{fig:ex100-discount-matrix}, while their overall effect on the consumption is displayed in Fig.~\ref{fig:ex100-consumption}. 
\newp{Visually, the individual discount matrices exhibit distinct patterns (cf. \texttt{Gurobi} and \texttt{Leap}), but the effective result stays comparable regarding the \coo\ reduction. The optimal \coo\ reduction is 12.45\,kg, \texttt{Leap} differs by 0.3\%, \texttt{Gurobi} by 0.6\%, \texttt{Decomp-Leap} by 1.4\% and \texttt{Decomp-Gurobi} by only $0.005\%$.}

\begin{figure}
    \centering
    \includegraphics[width=0.35\textwidth]{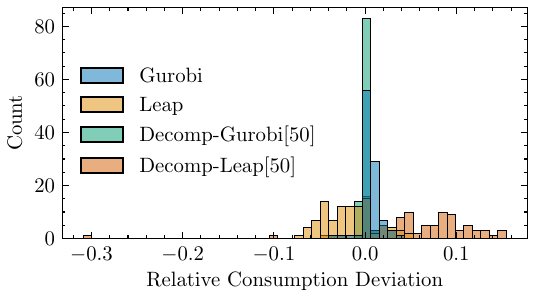}
    \caption{Histogram of the relative consumption deviation. One can see that both \texttt{Gurobi} solvers have relatively little spread. And are well centered around zero.
    The \texttt{Leap} solvers, on the other hand, possess a large spread and are additionally shifted away from zero. The shift away from zero reduces in larger problem instances.}
    \label{fig:ex100-consumption-deviation}
\end{figure}

\begin{figure}
    \centering
    \includegraphics[width=0.35\textwidth]{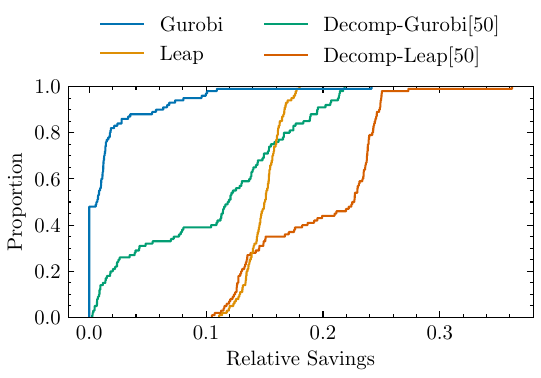}
    \caption{A cumulative distribution plot of the relative savings of the customers. The two chunks can be well distinguished in the \texttt{Decomp} solvers. \texttt{Gurobi} only distributes savings to relatively few customers. On the other hand, \texttt{Leap} distributes similar discounts to all customers.}
    \label{fig:ex100-fairness}
\end{figure}

Apart from the global optimization metrics, we are also interested in how the optimization performs per customer. In Fig.~\ref{fig:ex100-consumption-deviation}, one can see how the relative consumption changes are distributed. Furthermore, Fig.~\ref{fig:ex100-fairness} visualizes the distribution of savings to the customers.

Lastly, it remains important to note that the results for the \texttt{Leap} solvers vary throughout multiple runs. Here, only a single run has been picked, which is characteristic of the behavior of these solvers. Furthermore, no investigation towards direct QPU access has been made since the space requirements for a single customer are already 76 integer variables, i.e., 228 binary variables. The problem after gathering multiple customers in a chunk is, hence, not embeddable in the QPU since we are facing quite dense connectivity in the QUBO. For a reduced problem size, we perform investigations in Sec.~\ref{sec:eval-qpu}.

\subsection{Scaling Analysis}
\begin{figure}
    \centering
    \includegraphics[width=0.45\textwidth]{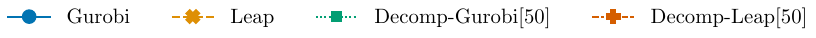}\\
    \includegraphics[width=0.45\textwidth]{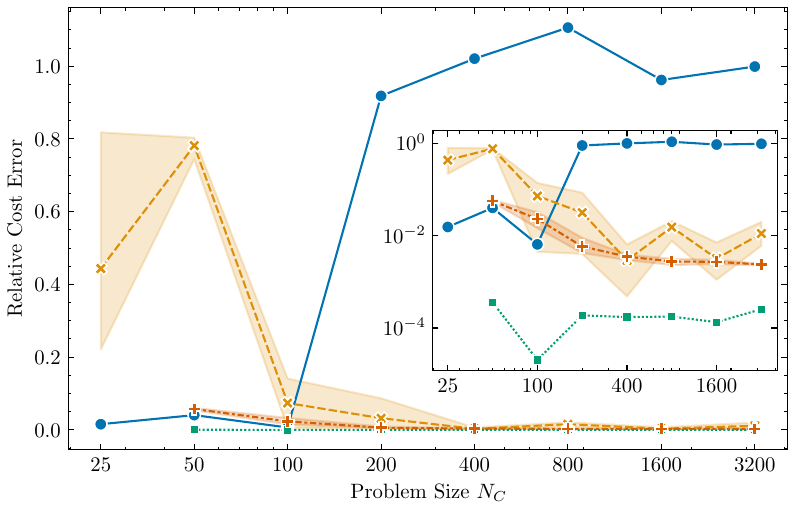}
    \caption{The relative cost error for different solvers with respect to problem size $N_C$. Cost is the optimization objective known from Eq.~\eqref{eq:objective}--\eqref{eq:discount_regularization}. The relative value is taken with respect to the bound known from $C(\zeta^*)$. The inset shows the relative cost error with logarithmic scaling. The error bands indicate the maximum and minimum of the three runs.}
    \label{fig:eval-full}
\end{figure}

\begin{figure}
    \centering
    \includegraphics[width=0.45\textwidth]{graphics/plots/full/legend.pdf}\\
    \includegraphics[width=0.24\textwidth,trim=0 15 0 0,clip]{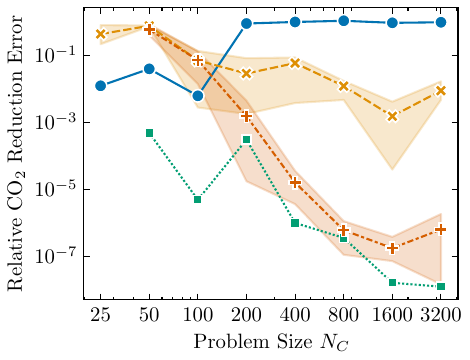}%
    \includegraphics[width=0.24\textwidth,trim=0 15 0 0,clip]{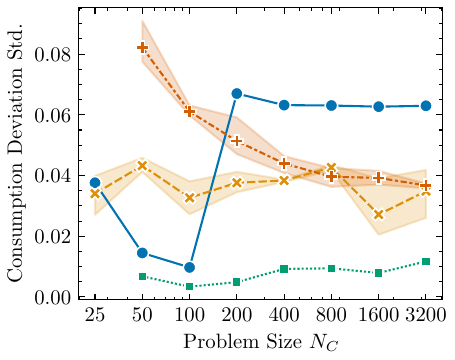}\\
    \includegraphics[width=0.24\textwidth]{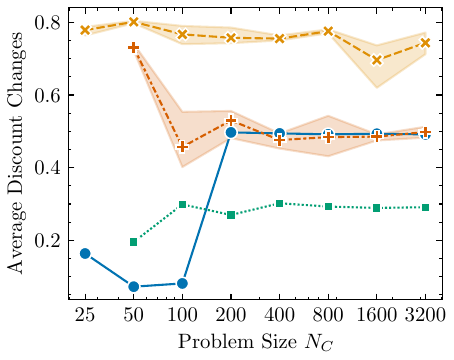}%
    \includegraphics[width=0.24\textwidth]{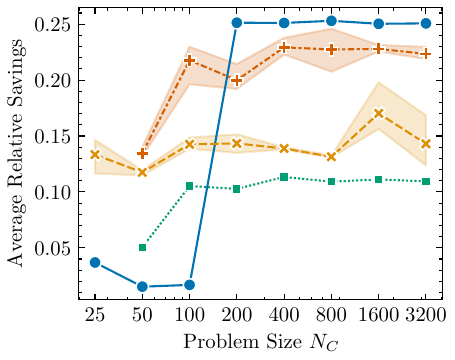}
    \caption{The auxiliary metrics for different problem sizes and different solvers. The plots show the \coo\ reduction error and the per-customer metrics: The standard deviation of all customer consumption deviations \eqref{eq:sconstr_consumption_deviation} the average discount changes \eqref{eq:discount_change_penalty}, which we both want to be small. The bottom right pane displays the average relative savings $\langle s_c \rangle$. The error bands indicate the maximum and minimum of the three runs.}
    \label{fig:eval-full2}
\end{figure}

To test the performance of different solvers, we created test instances using generated data with $N_C$ ranging from 25 to 3200 customers and considering the full 76 timesteps. Our problem instances, therefore, consist of 1,900 to 243,200 integer variables. 
To account for the stochasticity of the results from the quantum solvers, we run the quantum solvers three times.

\newp{The results in terms of the objective function depending on problem size are visualized in Fig.~\ref{fig:eval-full}. It is evident that a crossover in performance between \texttt{Gurobi} and \texttt{Leap} happens between 100 and 200 customers.} After that size, \texttt{Gurobi} \newp[cannot find converged results in the given time limit]{is not able to finish the root relaxation within the given time bounds and falls back to a heuristic solution, which has inferior performance}. Although not a directly fair comparison since \texttt{Gurobi} runs on a local machine while the \texttt{Leap} hybrid solver is run on a proprietary D-Wave cloud architecture, we argue that the pattern generalizes, i.e., the inflection point where \texttt{Gurobi} doesn't reach satisfactory results anymore shifts to larger instances but eventually happens. \newp{In the regime $N_C < 200$, Gurobi's MIP Gap roughly coincides with the relative cost error since the lower bound of Gurobi is almost equal to $C(\zeta^*)$}. For low problem size, the \texttt{Leap} solver demonstrates \newp{high relative cost error}, however, this error decreases rapidly up to low hundreds of consumers. Nonetheless, the decomposition routines outperform the general-purpose solvers, especially the purely classical \texttt{Decomp-Gurobi} approach. 

\newp{Fig.~\ref{fig:eval-full2} shows the relative \coo\ reduction error and three per-customer metrics. The relative \coo\ reduction error shows a similar pattern as the inset in Fig.~\ref{fig:eval-full}, which is due to the emission reduction being the main part of the optimization objective. It is apparent that the decomposition routines in the higher problem instances produce results with almost perfect \coo\ reduction (less than $10^-5$ error), which can be explained by the fact that they have access to the best emission reduction bound. As a consequence, the per-customer penalties (Eq.~\eqref{eq:sconstr_consumption_deviation}--Eq.~\eqref{eq:discount_regularization}) are responsible for the cost error visible in Fig.~\ref{fig:eval-full}.}

Investigating the per-customer constraints, we notice that the \texttt{Gurobi}-based solvers outperform the quantum-enhanced routines \newp{($N_C < 200$)}. This is likely due to \texttt{Gurobi} being better at handling smaller changes in the optimization objective. However, it is also important to note that, as apparent from the discount matrices in Fig.~\ref{fig:ex100-discount-matrix}, \texttt{Gurobi} gives many customers not even a single discount. Hence, they do not receive any discount changes or consumption deviations, which reduces the average measure. 

\newp{Examining the heuristic solutions of \texttt{Gurobi}, when solving its root relaxation aborts ($N_C \geq 200$), reveals that the discount matrix is almost completely filled with extremal values $z_{c,t} = \pm \zm$. The constraints are satisfied, but the discounts are randomly distributed, which allows for computing the per-customer metrics analytically, supposing $z_{c,t} = \pm \zm$ with equal probability. The consumption deviation metric from Eq.~\eqref{eq:consumption_deviation_metric} simplifies to
$\zm\sqrt{\langle D_c^{-2}\sum_t d^2_{c,t}\rangle_c}$, which is e.g. $0.064$ in the $N_C = 400$ case. The average discount change metric reduces to the probability of observing one discount change ($0.5$). Finally, $s_c = \zm^2 = 1/4$, since the customer savings are dependent on a weighted average over $z_{c,t}^2 = \zm^2 = \mathrm{const.}$, see Eq.~\eqref{eq:price_change}.}

To conclude this analysis, we remark that \texttt{Gurobi} \newp{struggles at large problem sizes since its root relaxation cannot be solved within the given time constraints}, which indicates a potential advantage of the quantum-enhanced solver here. Yet, the domain-specific decomposition routine provides even better results, especially in conjunction with the classical solver. We argue that since the decomposition-based solvers work so well, the space of good solutions is large, which makes this problem a fitting choice for heuristic-based solvers more than mathematical solvers, like \texttt{Gurobi}.

\subsection{Chunk Size Effect}
\begin{figure}
    \centering
    \includegraphics[width=0.35\textwidth]{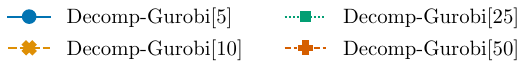}\\
    \includegraphics[width=0.24\textwidth]{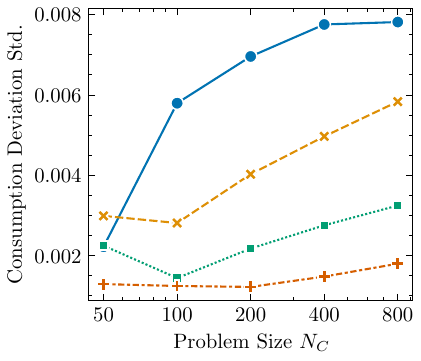}%
    \includegraphics[width=0.24\textwidth]{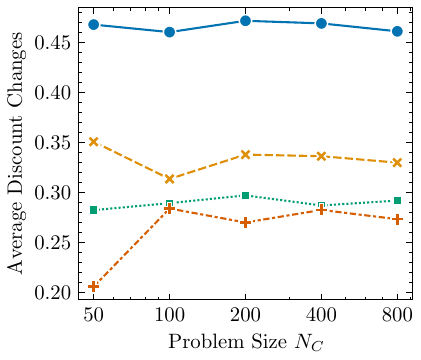}
    \caption{Per-customer metrics evaluated with different chunk sizes in the decomposition. As expected, the metrics improve (get smaller) as the chunk sizes get larger since more flexibility remains in the chunk.}
    \label{fig:eval-chunk-size}
\end{figure}

After we observed that the decomposition solver provides satisfying results both with \texttt{Gurobi} and \texttt{Leap} employed as sub-solver, we are interested in what impact the chunk size has on the result.
For that, we only inspect \texttt{Decomp-Gurobi} with different chunk sizes $m = 5, 10, 25, 50$ and focus on a reduced problem size frame up until $N_C = 800$. We have seen that the problem complexity does not grow linearly with the problem size. Thus, we give a more generous timeout of $0.5\,\mathrm{s} \times m$ in this investigation in order to isolate the effects of the decomposition routine from the solver performance%
\footnote{Preliminary experiments have shown a performance increase for the larger chunk sizes when increasing the timeout. This increase came to a slowdown at around $0.5\,\mathrm{s} \times m$.}.
The global effect, i.e., how much \coo\ is reduced, does not differ between the chunk sizes \newp{(below 1\% error)}. The constant sequential updating of the objective also helps a lot with finding the best \coo\ reduction, even with five customer chunks. Fig.~\ref{fig:eval-chunk-size} shows the \newp{consumption deviation and discount changes}, where a clear tendency that larger chunks result in lower per-customer metrics can be observed, i.e., less consumption deviation per customer and fewer overall discount changes.

\subsection{Fairness Analysis}

\begin{figure}
    \centering
    \includegraphics[width=0.35\textwidth]{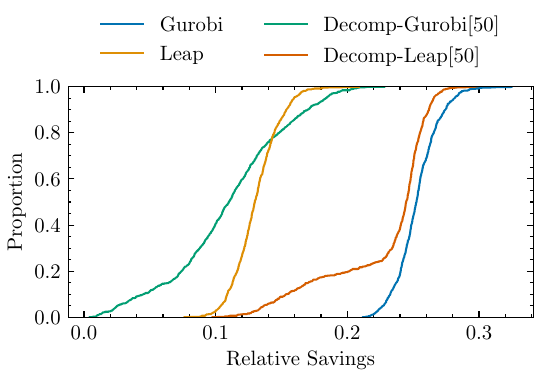}
    \caption{A cumulative distribution plot of the relative savings of the customers at $N_C = 800$. As discussed earlier, \texttt{Gurobi}s \newp{root relaxation does not finish anymore}, which causes savings of around 25\%. \texttt{Leap} produces fair discounts, similar to Fig.~\ref{fig:ex100-fairness}. The other two solvers produce more complex, unfair savings distributions.} 
    \label{fig:ex800-fairness}
\end{figure}

The goal of this section is to investigate how the solvers strategically distribute the discounts to the target customers. This is done by investigating how the relative savings $s_c$ are distributed between individual customers. Fig.~\ref{fig:ex100-fairness} and Fig.~\ref{fig:ex800-fairness} show two cumulative distribution plots of the results from 100 and 800 customer problem sizes. \newp{The more vertical (zero slope) a given cumulative line is, the fairer the discounts are distributed among the consumers, thereby implying a better social-welfare measure for the energy consumers.} Except for \texttt{Gurobi}, the qualitative patterns of the solvers are similar. \texttt{Leap} produces a fair savings distribution, which means that all customers experience the same savings \newp{(10\%--15\%)}. 

In Fig.~\ref{fig:ex100-fairness} the splitting in half of the decomposition can be observed quite remarkably. The resolution of the 16 individual chunks in Fig.~\ref{fig:ex800-fairness} is no longer possible. However, a kink in \texttt{Decomp-Leap} can be observed, which means that about 70\% of the customers save a similar and relatively large amount \newp{(22\%--25\%)}, while fewer savings are distributed to a smaller group \newp{(10\%--22\%)}. \texttt{Decomp-Gurboi} reveals a straight but shallow curve, which means that customers will receive savings between 0\% and 20\% almost equally likely.

\subsection{Direct QPU-Access with Decomposition}\label{sec:eval-qpu}

\begin{figure}
    \centering
    \includegraphics[height=1.5in]{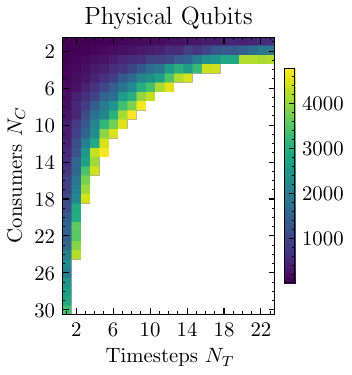}
    \includegraphics[height=1.5in]{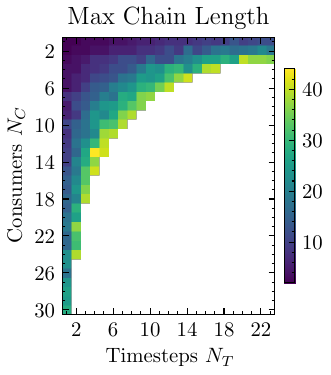}
    \caption{Embeddable sub-problem size for the D-Wave Advantage 4.1 QPU. The left-hand matrix shows how many physical qubits are needed when a sub-problem with $N_C$ customers and $N_T$ timesteps are embedded. A white field indicates that no embedding has been found. The right-hand plot shows the maximal chain length for the found embedding, i.e., how many qubits are maximally connected to form one logical qubit. All embeddings were found using D-Wave's \texttt{MinorMiner} package.}
    \label{fig:eval-embedding-size}
\end{figure}

A quantum annealing processor, such as D-Waves Advantage 4.1, suffers from limited connectivity between the physical qubits. However, for our QUBO sub-problems~\eqref{eq:sub_problem}, we can analytically compute the number of couplings for a single qubit as follows:
\newp{\begin{align}
    Q \left( N_T - 1\right) + Q \left( m - 1\right) + Q - 1,
\end{align}
where $Q$ is the number of qubits required to encode the discount, $N_T$ is the number of timesteps, and $m$ is the number of customers per chunk.} This term is derived by inspecting the terms in the QUBO formula and observing that we either have couplings within all customers of a chunk at a single timestep or couplings within all timesteps of a single customer. For the first case, one qubit is connected to all $Q$ qubits of the other $m - 1$ customers in the chunk and to $Q - 1$ qubits of the same customer. Analogue for the second case, but the $Q - 1$ connections within the timestep have already been covered in the first case.

The derived quantity grows with the problem size, but the couplings per qubit of the D-Waves Pegasus graph is a constant 15~\cite{mcgeoch}. Thus, physical qubits have to be chained together to logical qubits in order to allow for higher connectivity. Finding the best, so-called embedding, is itself an NP-hard optimization problem, for which we utilize D-Wave's heuristic \texttt{MinorMiner}.

\input{sections/qpu_evaluations/02_reduced}

%% file: sections/qpu_evaluations/02_reduced.tex
\begin{figure}
    \centering
    \includegraphics[width=0.45\textwidth]{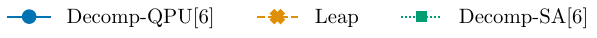}\\
    \includegraphics[width=0.24\textwidth,trim=0 15 0 0,clip]{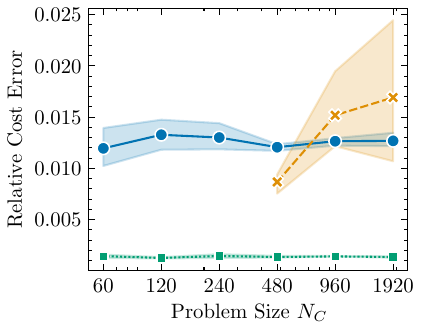}%
    \includegraphics[width=0.24\textwidth,trim=0 15 0 0,clip]{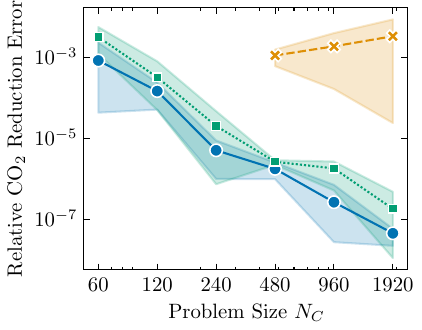}\\
    \includegraphics[width=0.24\textwidth]{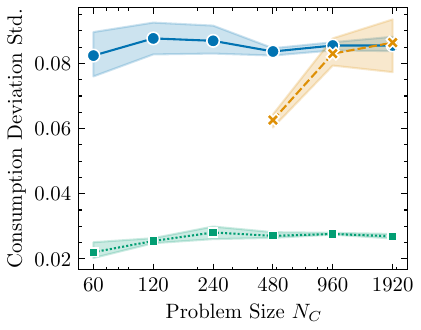}%
    \includegraphics[width=0.24\textwidth]{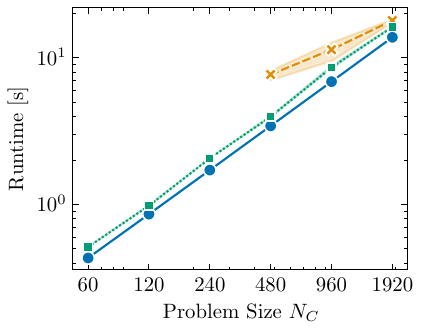}%
    \caption{All metrics for the small problem sizes, also considering QUBO sub-problems that can be solved using D-Wave's QPU. Comparing both Quantum routines, one can observe that \texttt{Decomp-QPU} returns results with slightly lower cost than \texttt{Leap}. Yet, the \coo\ reduction is vastly better in the decomposed case.
    Notably, the difference of the \texttt{Decomp} solvers in the relative cost error is mostly due to the higher consumption deviation from the QPU results. To match the runtime of the \texttt{Decomp-SA} to \texttt{Decomp-QPU}, we had to manually adjusted the number of samples from the simulated annealing routine to 30. We also observe that \texttt{Leap} outruns the set timeout here.}
    \label{fig:eval-qpu}
\end{figure}

Fig.~\ref{fig:eval-embedding-size} shows the computed embeddings for the sub-problem QUBOs with different problem sizes. It is apparent that we are very limited to small problem sizes. Since we do not want too few customers in a chunk to preserve flexibility, we settle at a reasonable middle ground of chunk size six and $12$ timesteps. We interpolate the original data to 12 timesteps and use various (multiples of 6) customer sizes to compare the performance of \texttt{Decomp-QPU} against the other solvers. For each sub-problem, we take 100 readings from the QPU. We cannot directly steer the timeout in this case. Thus, we first run \texttt{Decomp-QPU} and then set the timeout of the remaining solvers to exactly that time. However, \texttt{Leap} has a minimum runtime of 5\,s, which is the reason why we only include \texttt{Leap} in the cases where the \texttt{Decomp-QPU} time is more than 5\,s, being the case from $N_c = 480$ onwards. \newp{Embedding times are not considered since the embeddings are computed beforehand and remain constant for one timestep, chunk size ($N_T$, $m$) combination, independent of grid data.}

Again, we perform the analysis for different problem sizes, reaching from 60 to 1920 customers or 720 to 23,040 integer variables. The sub-problems comprise 72 integer variables, resulting in 216 binary variables in the QUBO formulation. In contrast to the previous analysis, we additionally investigate Simulated Annealing (SA) as a sub-problem QUBO solver in this instance. Due to the larger problem sizes of the previous sub-problems, the SA routine could not return results within the runtime boundaries we had set.
Fig.~\ref{fig:eval-qpu} displays the results of the experiments. The previously discussed solvers (\texttt{Gurobi}, \texttt{Decomp-Gurobi}, \texttt{Decomp-Leap}) exhibited similar patterns to the investigation done for the larger problem-sizes (Fig.~\ref{fig:eval-full}). Therefore, we only focus on the QPU and SA-based decomposition routines and \texttt{Leap}.

\newp{\texttt{Decomp-SA} exhibits the lowest cost error compared to the two quantum-enhanced methods}. The two \texttt{Decomp} solvers (one using classical and the other using quantum compute) demonstrate similar performance in terms of \coo\ reduction. Interestingly, the \newp{consumption deviation}, and therefore also the dominating factor in the cost, are measured at a very constant level between the problem sizes. Curiously, SA, as the sub-solver, performs better \newp{($\approx 2\%$)} concerning \newp{consumption deviation} than the QPU does \newp{($\approx 8\%$)}, leading to a gap in the cost. \texttt{Leap} exhibits similar performance as in our previous experiments. Most notably, although, is that \texttt{Decomp-QPU} seems to perform better than \texttt{Leap} regarding the optimization objective \newp{(1.3\% error versus 1.7\% error)}, and concerning the \coo\ reduction \newp{($4.5\times 10^{-8}$ error versus $3.4 \times 10^{-3}$ error)}. That indicates that our developed hybrid quantum routine does seem to outperform the general-purpose \texttt{Leap} for this particular task.

%% file: sections/06_conclusion.tex
\section{Conclusion}\label{sec:conclusion}

We explored the feasibility of current quantum computing techniques for DSR by developing a mathematical formulation that utilizes discount scheduling to shift grid load to more appropriate times. Our formulation involves providing discretized discounts to multiple customers at different times to incentivize a load shift while ensuring the total consumption stays fixed. We chose \coo\ emission reduction as the main objective for our DSP implementation of DSR. With secondary objectives, such as maintaining grid stability and ensuring customer well-being, we formulated a QIP problem.

Upon close inspection of the problem, we developed a custom decomposition algorithm that compartmentalizes the problem into customer chunks. These sub-problems involve unconstrained integer optimization and can be effectively addressed on quantum computers if encoded correctly. Moreover, since the problems are solved sequentially, we incorporated the accumulated errors into the subsequent optimization problems. Lastly, we developed a post-processing algorithm that further refines the solution.

In the end, we benchmarked the performance of a classical general-purpose solver against D-Wave's Leap hybrid quantum-classical solver and our customized decomposition method with various (quantum or classical) sub-solvers employed. We observed that the classical solver fails to produce acceptable results after a specific problem size when using a linearly increasing timeout for the problem size. In contrast, the quantum-enhanced \texttt{Leap} continues to provide satisfactory results. This indicates a potential advantage of solving this particular problem using \texttt{Leap} over the purely classical counterpart, \texttt{Gurobi}. Nonetheless, the decomposition method with the classical solver as sub-solver developed the best-achieved results over the range of problem sizes we investigated. Furthermore, using quantum or simulated annealing for the QUBO problems has resulted in good performance. We found that decomposition paired with quantum annealing returned comparable energies to \texttt{Leap}.

We remark that the pairing of the decomposition method with \texttt{Leap} with large chunk sizes might be a promising pathway for utilizing the quantum-enhanced method for huge instances of this problem. This statement requires further experiments, but we argue that solving large sub-problems within time constraints may pose challenges for \texttt{Gurobi}, whereas \texttt{Leap} could yield acceptable results. Further future work includes the response of the solvers to different problem parameter settings and making the grid constraints more physically realistic rather than our realistic yet pragmatic chosen constant band.

Lastly, determining precise energy requirements of quantum computing hardware and the trade-off between quantum computing algorithms runtime and used energy is an active and interesting area of research \cite{berger2021quantum, chen2023, qei}. If the community can demonstrate practical quantum advantage (e.g. quantum runtime of seconds or hours as opposed to weeks or years for classical HPC runtime) for use-cases which themselves reduce \coo\ emissions, perhaps this may offset the potential \coo\ cost resulting from the manufacturing or operating future quantum computers. We believe this work on how quantum computers may solve optimization problems related to the sustainable energy transition by embedding sustainability notions into the use-case itself is a progressive step towards using quantum computing for important global issues.

%

%% file: sections/a_glossary.tex
\section{Glossary}\label{sec:glossary}
For a better overview of the used symbols and parameters in the formulation of the DSP, we provide an overview in Table~\ref{tab:objects}.

\begin{table}
    \caption{The symbols and parameters in the DSP formulation}
    \begin{tabular}{cclc}
         Label & Range & Description & Units \\ \hline
         $c$ & $\set{1,\dots,N_C}$ & Customer index \\ 
         $t$ & $\set{1,\dots,N_T}$ & Timestep index \\
         $k$ & $\set{1,\dots,N_K}$ & Discount type index \\
         $d_{c,t}$ & $\mathbb{R}^+$ & Customer power draw & kWh \\
         $\chi_{c}$ & $\mathbb{R}^+$ & Customer price elasticity\\
         $I_t$ & $\mathbb{R}^+$ & Grid \coo\ intensity & $\frac{\mathrm{g}}{\mathrm{kWh}}$ \\
         $z_{c,t}$ & $Z$ & Discounts \\
         $\zm$ & $[0, 1]$ & Maximum discount &  \\ 
         $Z$ & $\{-\zm...\zm\}$ & Discount categories \\
         $Q$ & $\mathbb{N}$ & Number of bits in encoding \\
         $x_{c,t,k}$ & $\{0, 1\}$ & Binary decision variables \\
         $\Delta p_t$ & $\mathbb{R}^+$ & Momentary consumption & kWh \\ & & deviation bound  & \\
         $\lambda_1$ & $\mathbb{R}^+$ & Consumption deviation penalty &  \\
         $\lambda_2$ & $\mathbb{R}^+$ & Discount $L2$-regularization \\
         $\lambda_3$ & $\mathbb{R}^+$ & Discount change penalty\\
         $\zeta_t$ & $[-\zm, \zm]$ & Effective discount \\
         $D_c$ & $\mathbb{R}^+$ & Total consumption per customer & kWh \\
         $D_t$ & $\mathbb{R}^+$ & Total consumption per timestep & kWh \\
         $\widetilde D_t$ & $\mathbb{R}^+$ & Mutable consumption & kWh \\
         $M$ & $\mathbb{N}$ & Number of chunks\\
         $m$ & $\mathbb{N}$ & Chunk size\\
         $j$ & $\set{1,\dots,M}$ & Chunk index \\
         $\xi_t^j$ & $[-\zm, \zm]$ & Effective chunk discount \\
         $r$ & $\mathbb{N}$ & Post-processing cut-off
    \end{tabular}
    \label{tab:objects}
\end{table}

%

%% file: main.bbl
\begin{thebibliography}{10}
\providecommand{\url}[1]{#1}
\csname url@samestyle\endcsname
\providecommand{\newblock}{\relax}
\providecommand{\bibinfo}[2]{#2}
\providecommand{\BIBentrySTDinterwordspacing}{\spaceskip=0pt\relax}
\providecommand{\BIBentryALTinterwordstretchfactor}{4}
\providecommand{\BIBentryALTinterwordspacing}{\spaceskip=\fontdimen2\font plus
\BIBentryALTinterwordstretchfactor\fontdimen3\font minus \fontdimen4\font\relax}
\providecommand{\BIBforeignlanguage}[2]{{%
\expandafter\ifx\csname l@#1\endcsname\relax
\typeout{** WARNING: IEEEtran.bst: No hyphenation pattern has been}%
\typeout{** loaded for the language `#1'. Using the pattern for}%
\typeout{** the default language instead.}%
\else
\language=\csname l@#1\endcsname
\fi
#2}}
\providecommand{\BIBdecl}{\relax}
\BIBdecl

\bibitem{DSR_00}
J.~Torriti, \emph{\BIBforeignlanguage{eng}{Peak energy demand and demand side response}}, ser. Routledge explorations in environmental studies.\hskip 1em plus 0.5em minus 0.4em\relax London New York, NY: Earthscan from Routledge, 2016, {ISBN} 978-1-138-01625-5.

\bibitem{DSR_01}
P.~Siano, ``Demand response and smart grids—{A} survey,'' \emph{Renewable and Sustainable Energy Reviews}, vol.~30, pp. 461--478, Feb. 2014, \linkdoi{10.1016/j.rser.2013.10.022}.

\bibitem{DSR_02}
H.~T. Haider, O.~H. See \emph{et~al.}, ``A review of residential demand response of smart grid,'' \emph{Renewable and Sustainable Energy Reviews}, vol.~59, pp. 166--178, Jun. 2016, \linkdoi{10.1016/j.rser.2016.01.016}.

\bibitem{DSR_03}
N.~O'Connell, P.~Pinson \emph{et~al.}, ``Benefits and challenges of electrical demand response: {A} critical review,'' \emph{Renewable and Sustainable Energy Reviews}, vol.~39, pp. 686--699, Nov. 2014, \linkdoi{10.1016/j.rser.2014.07.098}.

\bibitem{DSR_04}
J.~S. Vardakas, N.~Zorba \emph{et~al.}, ``A {Survey} on {Demand} {Response} {Programs} in {Smart} {Grids}: {Pricing} {Methods} and {Optimization} {Algorithms},'' \emph{IEEE Communications Surveys \& Tutorials}, vol.~17, no.~1, pp. 152--178, 2015, \linkdoi{10.1109/COMST.2014.2341586}.

\bibitem{PhysRevLett.99.070502}
A.~Mizel, D.~A. Lidar \emph{et~al.}, ``Simple proof of equivalence between adiabatic quantum computation and the circuit model,'' \emph{Phys. Rev. Lett.}, vol.~99, p. 070502, Aug 2007, \linkdoi{10.1103/PhysRevLett.99.070502}.

\bibitem{RevModPhys.80.1061}
A.~Das and B.~K. Chakrabarti, ``Colloquium: Quantum annealing and analog quantum computation,'' \emph{Rev. Mod. Phys.}, vol.~80, pp. 1061--1081, Sep 2008, \linkdoi{10.1103/RevModPhys.80.1061}.

\bibitem{RevModPhys.90.015002}
T.~Albash and D.~A. Lidar, ``Adiabatic quantum computation,'' \emph{Rev. Mod. Phys.}, vol.~90, p. 015002, Jan 2018, \linkdoi{10.1103/RevModPhys.90.015002}.

\bibitem{Kadowaki:1998}
T.~Kadowaki and H.~Nishimori, ``Quantum annealing in the transverse ising model,'' \emph{Phys. Rev. E}, vol.~58, pp. 5355--5363, Nov 1998, \linkdoi{10.1103/PhysRevE.58.5355}.

\bibitem{FINNILA1994343}
A.~B. Finnila, M.~A. Gomez \emph{et~al.}, ``Quantum annealing: {A} new method for minimizing multidimensional functions,'' \emph{Chemical Physics Letters}, vol. 219, no.~5, pp. 343--348, Mar. 1994, \linkdoi{10.1016/0009-2614(94)00117-0}.

\bibitem{mcgeoch}
C.~McGeoch, P.~Farre \emph{et~al.}, ``\BIBforeignlanguage{en}{D-{Wave} {Hybrid} {Solver} {Service} + {Advantage}: {Technology} {Update}},'' \linkurl{https://www.dwavesys.com/media/m2xbmlhs/14-1048a-a\_d-wave\_hybrid\_solver\_service\_plus\_advantage\_technology\_update.pdf}.

\bibitem{QC_PA_01}
R.~Eskandarpour, K.~J. Bahadur~Ghosh \emph{et~al.}, ``Quantum-{Enhanced} {Grid} of the {Future}: {A} {Primer},'' \emph{IEEE Access}, vol.~8, pp. 188\,993--189\,002, 2020, \linkdoi{10.1109/ACCESS.2020.3031595}.

\bibitem{QC_PA_02}
Y.~Zhou, Z.~Tang \emph{et~al.}, ``Quantum computing in power systems,'' \emph{iEnergy}, vol.~1, no.~2, pp. 170--187, Jun. 2022, \linkdoi{10.23919/IEN.2022.0021}.

\bibitem{QC_PA_03}
D.~Tylavsky and G.~Heydt, ``Quantum computing in power system simulation,'' in \emph{2003 {IEEE} {Power} {Engineering} {Society} {General} {Meeting} ({IEEE} {Cat}. {No}.{03CH37491})}, vol.~2, Jul. 2003, pp. 950--956 Vol. 2, \linkdoi{10.1109/PES.2003.1270438}.

\bibitem{QC_PA_04}
R.~Eskandarpour, A.~Khodaei \emph{et~al.}, ``Quantum computing applications in power systems,'' \emph{CIGRE US National Committee, 2019 Grid of the Future Symposium}, 2019.

\bibitem{QC_PF01}
R.~Eskandarpour, K.~Ghosh \emph{et~al.}, ``Experimental {Quantum} {Computing} to {Solve} {Network} {DC} {Power} {Flow} {Problem},'' Jun. 2021, arXiv:2106.12032 [quant-ph]. \linkdoi{10.48550/arXiv.2106.12032}.

\bibitem{QC_PF_02}
R.~Eskandarpour, K.~Ghosh \emph{et~al.}, ``Quantum {Computing} {Solution} of {DC} {Power} {Flow},'' Oct. 2020, arXiv:2010.02442 [quant-ph]. \linkdoi{10.48550/arXiv.2010.02442}.

\bibitem{QC_cluster}
S.~DiAdamo, C.~O’Meara \emph{et~al.}, ``Practical {Quantum} {K}-{Means} {Clustering}: {Performance} {Analysis} and {Applications} in {Energy} {Grid} {Classification},'' \emph{IEEE Transactions on Quantum Engineering}, vol.~3, pp. 1--16, 2022, \linkdoi{10.1109/TQE.2022.3185505}.

\bibitem{dwave-speedup}
V.~S. Denchev, S.~Boixo \emph{et~al.}, ``What is the {Computational} {Value} of {Finite}-{Range} {Tunneling}?'' \emph{Physical Review X}, vol.~6, no.~3, p. 031015, Aug. 2016, \linkdoi{10.1103/PhysRevX.6.031015}.

\bibitem{lucas2014}
A.~Lucas, ``Ising formulations of many {NP} problems,'' \emph{Frontiers in Physics}, vol.~2, 2014, \linkdoi{10.3389/fphy.2014.00005}.

\bibitem{farhi2014a}
E.~Farhi, J.~Goldstone \emph{et~al.}, ``\BIBforeignlanguage{en}{A {Quantum} {Approximate} {Optimization} {Algorithm}},'' Nov. 2014, arXiv:1411.4028 [quant-ph]. \linkdoi{10.48550/arXiv.1411.4028}.

\bibitem{QC_O_02}
N.~Nikmehr, P.~Zhang \emph{et~al.}, ``Quantum {Distributed} {Unit} {Commitment}: {An} {Application} in {Microgrids},'' \emph{IEEE Transactions on Power Systems}, vol.~37, no.~5, pp. 3592--3603, Sep. 2022, \linkdoi{10.1109/TPWRS.2022.3141794}.

\bibitem{QC_O_03}
N.~Nikmehr, P.~Zhang \emph{et~al.}, ``Quantum-{Enabled} {Distributed} {Unit} {Commitment},'' in \emph{2022 {IEEE} {Power} \& {Energy} {Society} {General} {Meeting} ({PESGM})}, Jul. 2022, pp. 01--05, \linkdoi{10.1109/PESGM48719.2022.9917029}.

\bibitem{QC_O_04}
S.~Koretsky, P.~Gokhale \emph{et~al.}, ``Adapting {Quantum} {Approximation} {Optimization} {Algorithm} ({QAOA}) for {Unit} {Commitment},'' in \emph{2021 {IEEE} {International} {Conference} on {Quantum} {Computing} and {Engineering} ({QCE})}, Oct. 2021, pp. 181--187, \linkdoi{10.1109/QCE52317.2021.00035}.

\bibitem{QC_O_05}
F.~Feng, P.~Zhang \emph{et~al.}, ``Novel {Resolution} of {Unit} {Commitment} {Problems} {Through} {Quantum} {Surrogate} {Lagrangian} {Relaxation},'' \emph{IEEE Transactions on Power Systems}, vol.~38, no.~3, pp. 2460--2471, May 2023, \linkdoi{10.1109/TPWRS.2022.3181221}.

\bibitem{QC_O_01}
A.~Ajagekar and F.~You, ``Quantum computing for energy systems optimization: {Challenges} and opportunities,'' \emph{Energy}, vol. 179, pp. 76--89, Jul. 2019, \linkdoi{10.1016/j.energy.2019.04.186}.

\bibitem{Qc_ADMM}
C.~Gambella and A.~Simonetto, ``Multiblock {ADMM} heuristics for mixed-binary optimization on classical and quantum computers,'' \emph{{IEEE} Transactions on Quantum Engineering}, vol.~1, pp. 1--22, 2020, \linkdoi{10.1109/tqe.2020.3033139}.

\bibitem{Qc_Benders}
Z.~Zhao, L.~Fan \emph{et~al.}, ``Hybrid {Quantum} {Benders} {Decomposition} {For} {Mixed}-integer {Linear} {Programming},'' in \emph{2022 {IEEE} {Wireless} {Communications} and {Networking} {Conference} ({WCNC})}.\hskip 1em plus 0.5em minus 0.4em\relax Austin, TX, USA: IEEE Press, Apr. 2022, pp. 2536--2540, \linkdoi{10.1109/WCNC51071.2022.9771632}.

\bibitem{QC_O_09}
C.~O’Meara, M.~Fernández-Campoamor \emph{et~al.}, ``Quantum software architecture blueprints for the cloud: Overview and application to peer-2-peer energy trading,'' in \emph{2023 IEEE Conference on Technologies for Sustainability (SusTech)}, 2023, pp. 191--198, \linkdoi{10.1109/SusTech57309.2023.10129617}.

\bibitem{QC_O_08}
M.~Fernández-Campoamor, C.~O'Meara \emph{et~al.}, ``Community {Detection} in {Electrical} {Grids} {Using} {Quantum} {Annealing},'' Dec. 2021, arXiv:2112.08300 [quant-ph]. \linkdoi{10.48550/arXiv.2112.08300}.

\bibitem{QC_coalition}
S.~M. Venkatesh, A.~Macaluso \emph{et~al.}, ``{BILP}-{Q}: quantum coalition structure generation,'' in \emph{Proceedings of the 19th {ACM} {International} {Conference} on {Computing} {Frontiers}}, ser. {CF} '22.\hskip 1em plus 0.5em minus 0.4em\relax New York, NY, USA: Association for Computing Machinery, May 2022, pp. 189--192, \linkdoi{10.1145/3528416.3530235}.

\bibitem{QC_coalition_2}
S.~M. Venkatesh, A.~Macaluso \emph{et~al.}, ``{GCS}-{Q}: {Quantum} {Graph} {Coalition} {Structure} {Generation},'' in \emph{Computational {Science} – {ICCS} 2023: 23rd {International} {Conference}, {Prague}, {Czech} {Republic}, {July} 3–5, 2023, {Proceedings}, {Part} {V}}.\hskip 1em plus 0.5em minus 0.4em\relax Berlin, Heidelberg: Springer-Verlag, Jul. 2023, pp. 138--152, \linkdoi{10.1007/978-3-031-36030-5\_11}.

\bibitem{leib2023}
D.~Leib, T.~Seidel \emph{et~al.}, ``\BIBforeignlanguage{en}{An optimization case study for solving a transport robot scheduling problem on quantum-hybrid and quantum-inspired hardware},'' \emph{\BIBforeignlanguage{en}{Scientific Reports}}, vol.~13, no.~1, p. 18743, Oct. 2023, \linkdoi{10.1038/s41598-023-45668-1}.

\bibitem{geitz2022}
M.~Geitz, C.~Grozea \emph{et~al.}, ``\BIBforeignlanguage{en}{Solving the {Extended} {Job} {Shop} {Scheduling} {Problem} with {AGVs} – {Classical} and {Quantum} {Approaches}},'' in \emph{\BIBforeignlanguage{en}{Integration of {Constraint} {Programming}, {Artificial} {Intelligence}, and {Operations} {Research}}}, P.~Schaus, Ed.\hskip 1em plus 0.5em minus 0.4em\relax Springer International Publishing, 2022, pp. 120--137, \linkdoi{10.1007/978-3-031-08011-1\_10}.

\bibitem{colucci2023}
G.~Colucci, S.~v.~d. Linde \emph{et~al.}, ``Power {Network} {Optimization}: {A} {Quantum} {Approach},'' \emph{IEEE Access}, vol.~11, pp. 98\,926--98\,938, 2023, \linkdoi{10.1109/ACCESS.2023.3312997}.

\bibitem{oshiyama2022}
H.~Oshiyama and M.~Ohzeki, ``\BIBforeignlanguage{en}{Benchmark of quantum-inspired heuristic solvers for quadratic unconstrained binary optimization},'' \emph{\BIBforeignlanguage{en}{Scientific Reports}}, vol.~12, no.~1, p. 2146, Feb. 2022, \linkdoi{10.1038/s41598-022-06070-5}.

\bibitem{chapuis2017}
G.~Chapuis, H.~Djidjev \emph{et~al.}, ``Finding {Maximum} {Cliques} on a {Quantum} {Annealer},'' in \emph{Proceedings of the {Computing} {Frontiers} {Conference}}, ser. {CF}'17.\hskip 1em plus 0.5em minus 0.4em\relax New York, NY, USA: Association for Computing Machinery, May 2017, pp. 63--70, \linkdoi{10.1145/3075564.3075575}.

\bibitem{hahn2017}
G.~Hahn and H.~Djidjev, ``Reducing {Binary} {Quadratic} {Forms} for {More} {Scalable} {Quantum} {Annealing},'' in \emph{2017 {IEEE} {International} {Conference} on {Rebooting} {Computing} ({ICRC})}, Nov. 2017, pp. 1--8, \linkdoi{10.1109/ICRC.2017.8123654}.

\bibitem{pelofske2019}
E.~Pelofske, G.~Hahn \emph{et~al.}, ``\BIBforeignlanguage{en}{Solving {Large} {Maximum} {Clique} {Problems} on a {Quantum} {Annealer}},'' in \emph{\BIBforeignlanguage{en}{Quantum {Technology} and {Optimization} {Problems}}}, ser. Lecture {Notes} in {Computer} {Science}, S.~Feld and C.~Linnhoff-Popien, Eds.\hskip 1em plus 0.5em minus 0.4em\relax Cham: Springer International Publishing, 2019, pp. 123--135, \linkdoi{10.1007/978-3-030-14082-3\_11}.

\bibitem{pelofske2020}
E.~Pelofske, G.~Hahn \emph{et~al.}, ``\BIBforeignlanguage{en}{Decomposition {Algorithms} for {Solving} {NP}-hard {Problems} on a {Quantum} {Annealer}},'' \emph{\BIBforeignlanguage{en}{Journal of Signal Processing Systems}}, vol.~93, no.~4, pp. 405--420, Apr. 2021, \linkdoi{10.1007/s11265-020-01550-1}.

\bibitem{pelofske2019a}
E.~Pelofske, G.~Hahn \emph{et~al.}, ``Solving large minimum vertex cover problems on a quantum annealer,'' in \emph{Proceedings of the 16th {ACM} {International} {Conference} on {Computing} {Frontiers}}, ser. {CF} '19.\hskip 1em plus 0.5em minus 0.4em\relax New York, NY, USA: Association for Computing Machinery, Apr. 2019, pp. 76--84, \linkdoi{10.1145/3310273.3321562}.

\bibitem{guerreschi2021}
G.~G. Guerreschi, ``Solving {Quadratic} {Unconstrained} {Binary} {Optimization} with divide-and-conquer and quantum algorithms,'' Jan. 2021, arXiv:2101.07813 [quant-ph]. \linkdoi{10.48550/arXiv.2101.07813}.

\bibitem{zhou2023}
Z.~Zhou, Y.~Du \emph{et~al.}, ``{QAOA}-in-{QAOA}: {Solving} {Large}-{Scale} {MaxCut} {Problems} on {Small} {Quantum} {Machines},'' \emph{Physical Review Applied}, vol.~19, no.~2, p. 024027, Feb. 2023, \linkdoi{10.1103/PhysRevApplied.19.024027}.

\bibitem{tomesh2022a}
T.~Tomesh, Z.~H. Saleem \emph{et~al.}, ``\BIBforeignlanguage{en}{Quantum {Local} {Search} with the {Quantum} {Alternating} {Operator} {Ansatz}},'' \emph{\BIBforeignlanguage{en}{Quantum}}, vol.~6, p. 781, Aug. 2022, arXiv:2107.04109 [quant-ph]. \linkdoi{10.22331/q-2022-08-22-781}.

\bibitem{saleem2022}
Z.~H. Saleem, T.~Tomesh \emph{et~al.}, ``\BIBforeignlanguage{en}{Divide and {Conquer} for {Combinatorial} {Optimization} and {Distributed} {Quantum} {Computation}},'' Jul. 2022, arXiv:2107.07532 [quant-ph]. \linkdoi{10.48550/arXiv.2107.07532}.

\bibitem{fujii2022a}
K.~Fujii, K.~Mizuta \emph{et~al.}, ``Deep {Variational} {Quantum} {Eigensolver}: {A} {Divide}-{And}-{Conquer} {Method} for {Solving} a {Larger} {Problem} with {Smaller} {Size} {Quantum} {Computers},'' \emph{PRX Quantum}, vol.~3, no.~1, p. 010346, Mar. 2022, \linkdoi{10.1103/PRXQuantum.3.010346}.

\bibitem{booth2017a}
M.~Booth, S.~P. Reinhardt \emph{et~al.}, ``\BIBforeignlanguage{en}{Partitioning {Optimization} {Problems} for {Hybrid} {Classical}/{Quantum} {Execution}},'' \emph{\BIBforeignlanguage{en}{D-Wave Technical Report Series}}, Oct. 2017, \linkurl{https://docs.ocean.dwavesys.com/projects/qbsolv/en/latest/_downloads/bd15a2d8f32e587e9e5997ce9d5512cc/qbsolv_techReport.pdf}.

\bibitem{born1928}
M.~Born and V.~Fock, ``\BIBforeignlanguage{de}{Beweis des {Adiabatensatzes}},'' \emph{\BIBforeignlanguage{de}{Zeitschrift für Physik}}, vol.~51, no.~3, pp. 165--180, Mar. 1928, \linkdoi{10.1007/BF01343193}.

\bibitem{johnson2011}
M.~W. Johnson, M.~H.~S. Amin \emph{et~al.}, ``\BIBforeignlanguage{en}{Quantum annealing with manufactured spins},'' \emph{\BIBforeignlanguage{en}{Nature}}, vol. 473, no. 7346, pp. 194--198, May 2011, \linkdoi{10.1038/nature10012}.

\bibitem{alberini2011}
A.~Alberini, W.~Gans \emph{et~al.}, ``Residential consumption of gas and electricity in the {U}.{S}.: {The} role of prices and income,'' \emph{Energy Economics}, vol.~33, no.~5, pp. 870--881, Sep. 2011, \linkdoi{10.1016/j.eneco.2011.01.015}.

\bibitem{filippini2011}
M.~Filippini, ``Short- and long-run time-of-use price elasticities in {Swiss} residential electricity demand,'' \emph{Energy Policy}, vol.~39, no.~10, pp. 5811--5817, Oct. 2011, \linkdoi{10.1016/j.enpol.2011.06.002}.

\bibitem{dutta2017}
G.~Dutta and K.~Mitra, ``\BIBforeignlanguage{en}{A literature review on dynamic pricing of electricity},'' \emph{\BIBforeignlanguage{en}{Journal of the Operational Research Society}}, vol.~68, no.~10, pp. 1131--1145, Oct. 2017, \linkdoi{10.1057/s41274-016-0149-4}.

\bibitem{nelder1965}
J.~A. Nelder and R.~Mead, ``A {Simplex} {Method} for {Function} {Minimization},'' \emph{The Computer Journal}, vol.~7, no.~4, pp. 308--313, Jan. 1965, \linkdoi{10.1093/comjnl/7.4.308}.

\bibitem{gurobi2023}
{Gurobi Optimization, LLC}, ``{Gurobi Optimizer Reference Manual},'' 2023, \linkurl{https://www.gurobi.com}.

\bibitem{glover1986}
F.~Glover, ``Future paths for integer programming and links to artificial intelligence,'' \emph{Computers \& Operations Research}, vol.~13, no.~5, pp. 533--549, Jan. 1986, \linkdoi{10.1016/0305-0548(86)90048-1}.

\bibitem{berger2021quantum}
C.~Berger, A.~Di~Paolo \emph{et~al.}, ``Quantum technologies for climate change: {Preliminary} assessment,'' Jun. 2021, arXiv:2107.05362 [quant-ph]. \linkdoi{10.48550/arXiv.2107.05362}.

\bibitem{chen2023}
S.~Chen, ``Are quantum computers really energy efficient?'' \emph{Nature Computational Science}, vol.~3, no.~6, pp. 457--460, Jun. 2023, \linkdoi{10.1038/s43588-023-00459-6}.

\bibitem{qei}
``{QEI} 2023 – {Quantum} {Energy} {Initiative},'' \linkurl{https://quantum-energy-initiative.org/qei2023/}.

\end{thebibliography}
